\documentclass[prd,nofootinbib,superscriptaddress,preprint]{revtex4}
\usepackage[T1]{fontenc}
\usepackage[utf8]{inputenc} 
\usepackage{amsmath,amssymb}
\usepackage{epsfig}
\usepackage{graphicx,array}
\usepackage[usenames,dvipsnames]{color}
\usepackage{slashed}
\usepackage{comment}
\usepackage{booktabs}

\usepackage[colorlinks,citecolor=blue]{hyperref}
\usepackage{pdfpages}
\usepackage{color}
\usepackage{subfigure}

\usepackage{multirow}

\begin{document}

\title{Neutrino texture-zeros after JUNO's first results: \\ Implications for long-baseline neutrino experiments} 
	\author{Debasish Borah}
	\email{dborah@iitg.ac.in}
	\affiliation{Department of Physics, Indian Institute of Technology, Guwahati, Assam 781039, India}

        \author{Pritam Das}
	\email{prtmdas9@gmail.com}
 \affiliation{Department of Physics, Salbari College, Baksa, Assam 781318, India}
 
    \author{Debajyoti Dutta}
    \email{phy.debajyoti@bhattadevuniversity.ac.in}
    \affiliation{Department of Physics, Bhattadev University, Bajali, Pathsala, Assam 781325, India}

\begin{abstract}
The recent results from the JUNO reactor neutrino experiment have significantly improved our knowledge of the solar mixing angle $\theta_{12}$ and the solar mass splitting $\Delta m^2_{21}$. We study the impact of these improved estimates on the validity of texture-zeros in the light neutrino mass matrix by assuming neutrinos to be of Majorana nature. Considering a diagonal charged lepton basis, we revisit the previously allowed one-zero and two-zero textures and check their validity by using updated neutrino data from JUNO. While JUNO data rule out one previously allowed two-zero texture, they also make predictions for other neutrino parameters more precise. We finally study the prospects of probing the currently allowed texture-zeros and their predicted correlations among neutrino parameters at the Deep Underground Neutrino Experiment (DUNE). The inclusion of JUNO and reactor experiments strengthens DUNE's ability to constrain the allowed parameter space of both one-zero and two-zero textures. We also observe that DUNE benefits substantially from the complementarity with the T2HK experiment.
\end{abstract}

\maketitle 

\section{Introduction}\label{sec1}
The JUNO experiment has recently released its first results which constrain the solar mixing angle and solar mass-squared difference as \cite{JUNO:2025gmd}
\begin{equation}
    \sin^2{\theta_{12}} = 0.3092 \pm 0.0087, \,\,\,\, \Delta m^2_{21}= (7.50 \pm 0.12) \times 10^{-5} \, {\rm eV}^2
    \label{bfjuno}
 \end{equation}
at $1\sigma$ confidence level (CL). This is a significant improvement of precision by a factor of $1.6$ relative to the combination of all previous measurements. This improved measurement has interesting implications for a variety of neutrino physics scenarios and has already been studied in several works related to eV sterile neutrinos \cite{Minakata:2025azk}, non-unitarity \cite{Xing:2025bdm, Huang:2025znh}, dark matter \cite{Chao:2025sao}, leptonic flavor symmetries \cite{Zhang:2025jnn, Ge:2025csr, Jiang:2025hvq, He:2025idv, Petcov:2025aci, Ding:2025dzc, Ding:2025dqd}, neutrinoless double beta decay ($0\nu \beta \beta$) \cite{Ge:2025cky}, astrophysical neutrinos \cite{Xing:2025xte}, grand unified theories (GUT) \cite{Chen:2025afg, Saad:2025dkh}, Lorentz violation \cite{Araya-Santander:2025jfd}, global fits \cite{Capozzi:2025ovi, Esteban:2026phq} as well as synergies with other neutrino experiments \cite{Goswami:2025wla}.

In this work, we revisit the texture-zeros of the neutrino mass matrix and perform a comparative study of their validity in view of the neutrino global fit data ({\tt NuFit 6.0}) \cite{Esteban:2024eli} and JUNO's results \cite{JUNO:2025gmd}. Texture-zeros in the neutrino mass matrix has been studied extensively in the literature, a review of which can be found in \cite{Ludl:2014axa}. The predictive nature of such texture-zero scenarios makes them testable in a variety of experiments, ranging from neutrino oscillations, $0\nu \beta \beta$ as well as cosmic microwave background (CMB) missions like PLANCK \cite{Planck:2018vyg}. Such observables or experimental data already rule out more than two texture-zeros in the neutrino mass matrix in the diagonal charged lepton basis. A non-exhaustive list of phenomenological studies of such one-zero and two-zero textures can be found in \cite{Xing:2003ic, Lashin:2011dn, Deepthi:2011sk, Gautam:2015kya, Cebola:2015dwa, Frampton:2002yf, Xing:2002ta, Xing:2002ap, Kageyama:2002zw, Dev:2006qe, Ludl:2011vv, Kumar:2011vf, Fritzsch:2011qv, Meloni:2012sx, Meloni:2014yea, Dev:2014dla, Dev:2015lya, Borah:2015vra, Kaneko:2002yp, Kaneko:2003cy, Dev:2010vy, Bando:2004hi, Nguyen:2014mwa, Kalita:2015tda, Bora:2016ygl, Kitabayashi:2018bye, Borgohain:2019pya, Kitabayashi:2019uzg, Borgohain:2020now, Kitabayashi:2020ajn, Borgohain:2020csn, Gautam:2021vhf, Hyodo:2021kqf, Minamizawa:2022fch, Hyodo:2022xyn, Calibbi:2025ded,Denton:2023hkx}. While we adopt a model-independent approach in this work, several earlier works studied flavour symmetry-based origins of such texture-zeros in the neutrino mass matrix \cite{Berger:2000zj, Low:2004wx, Low:2005yc, Grimus:2004hf, Xing:2009hx, Dev:2011jc, Araki:2012ip, GonzalezFelipe:2014zjk, Dighe:2009xj, Grimus:2004az}.

We first check the validity of previously allowed one-zero and two-zero textures in light of the latest neutrino global fit data \cite{Esteban:2024eli} and JUNO's results \cite{JUNO:2025gmd} by assuming neutrinos to be of Majorana nature and the charged lepton mass matrix to be diagonal. For the texture-zeros allowed by neutrino oscillation data, we check their validity in view of stringent lower bounds on $0\nu\beta \beta$ half-life \cite{KamLAND-Zen:2024eml} and cosmological upper bound on sum of absolute neutrino masses \cite{Planck:2018vyg}. While JUNO's results affect the one-zero textures marginally, they rule out one of the previously allowed two-zero textures. The second part of the work involves the study of these currently allowed texture-zeros from all experimental constraints in the context of the upcoming Deep Underground Neutrino Experiment (DUNE) \cite{DUNE:2015lol,DUNE:2015lol, DUNE:2020ypp} and Tokai to Hyper-Kamiokande (T2HK) \cite{Hyper-KamiokandeProto-:2015xww}. Using the projected future sensitivity of DUNE, we investigate how effectively these texture-based neutrino mass models can be constrained further. In particular, we study the extent to which DUNE can test and restrict the allowed parameter space of these textures after a given exposure. We use both {\tt NuFit 6.0} as well as JUNO's first results to analyze the prospects of constraining these texture-zeros by combining DUNE and T2HK \cite{Hyper-KamiokandeProto-:2015xww}. 

This paper is organized as follows. In section \ref{sec2} we list out the texture-zero mass matrices. In section \ref{sec3}, we discuss our numerical analysis and the corresponding results based on the latest neutrino global fit data and JUNO's first results. In section \ref{sec4}, we study the prospects of discriminating the currently allowed texture-zeros at future long-baseline neutrino experiments. Finally, we conclude in section \ref{sec5}.

\section{Texture-Zeros}
\label{sec2}
The fact that neutrinos have tiny but non-zero masses and large mixing has already been established \cite{ParticleDataGroup:2024cfk}. While the nature of neutrinos is not known yet, the possibility of neutrinos being Majorana fermions has been studied extensively in the literature due to their minimal nature. For the Majorana nature of light neutrinos, the corresponding $3\times 3$ mass matrix $M_{\nu}$ remains complex symmetric with six independent complex parameters. 
Therefore, the number of possible mass matrices with $k$ texture-zeros is given by 
\begin{equation}
\label{prmtn}
^6C_k=\frac{6!}{k!(6-k)!}
\end{equation}
A symmetric mass matrix with more than three texture zeros is not compatible with the observed leptonic mixing and masses. The Pontecorvo-Maki-Nakagawa-Sakata (PMNS) leptonic mixing matrix is related to the diagonalizing matrices of charged lepton and neutrino mass matrices as 
\begin{equation}
U_{\text{PMNS}} = U^{\dagger}_l U_{\nu}
\label{pmns0}
\end{equation}
where $U_l, U_\nu$ denote the diagonalizing matrices of charged lepton and neutrino mass matrices respectively.
In the diagonal charged lepton basis, $U_{\text{PMNS}}$ is the same as the diagonalising matrix $U_{\nu}$ of the neutrino mass matrix $M_{\nu}$. In this basis, it has already been established \cite{Xing:2004ik} that a symmetric Majorana neutrino mass matrix with 3 texture zeros is not compatible with neutrino oscillation data. This leaves us with the possible one-zero and two-zero texture mass matrices. Going by the counting formula in Eq. \eqref{prmtn}, we can have six possible one-zero texture and fifteen possible two-zero texture mass matrices. The one-zero texture mass matrices can be written as 
\begin{center}

$ G_1 :\left(\begin{array}{ccc}
0& \times&\times\\
\times& \times&\times \\
\times& \times&\times 
\end{array}\right) , G_2 :\left(\begin{array}{ccc}
\times& 0&\times\\
0& \times&\times \\
\times& \times&\times 
\end{array}\right) , G_3 :\left(\begin{array}{ccc}
\times& \times&0\\
\times& \times&\times \\
0 & \times&\times 
\end{array}\right) ,   G_4 :\left(\begin{array}{ccc}
\times& \times&\times\\
\times & 0 &\times \\
\times& \times&\times 
\end{array}\right)  ,$
 
\end{center}
\begin{equation}
G_5 :\left(\begin{array}{ccc}
\times& \times&\times\\
\times& \times & 0 \\
\times& 0 &\times 
\end{array}\right) , 
 G_6 :\left(\begin{array}{ccc}
\times& \times&\times\\
\times& \times& \times \\
\times& \times & 0
\end{array}\right), 
\end{equation}
where $\times$ denotes non-zero arbitrary elements of $M_{\nu}$. Out of the fifteen two-zero textures, nine were already out by neutrino oscillation and cosmology data \cite{Fritzsch:2011qv, Meloni:2014yea}. However, since there are ways to relax cosmological bounds on neutrino mass \cite{Naredo-Tuero:2024sgf, Shao:2024mag}, we consider all the two-zero textures which are allowed by neutrino data prior to JUNO's results. This leads to seven possible two-zero textures which can be listed, using the notations of \cite{Fritzsch:2011qv}, as
\begin{equation}
A_1 :\left(\begin{array}{ccc}
0& 0&\times\\
0& \times&\times \\
\times& \times&\times 
\end{array}\right) , 
 A_2 :\left(\begin{array}{ccc}
0& \times&0\\
\times& \times&\times \\
0& \times&\times 
\end{array}\right);
\end{equation}

\begin{equation}
B_1 :\left(\begin{array}{ccc}
\times& \times&0\\
\times& 0&\times \\
0& \times&\times 
\end{array}\right) , 
 B_2 :\left(\begin{array}{ccc}
\times& 0 &\times\\
0& \times&\times \\
\times& \times&0 
\end{array}\right),
 B_3 :\left(\begin{array}{ccc}
\times& 0&\times\\
0& 0&\times \\
\times& \times&\times 
\end{array}\right), 
B_4 :\left(\begin{array}{ccc}
\times& \times&0\\
\times& \times&\times \\
0& \times&0 
\end{array}\right);
\end{equation}

\begin{equation}
C :\left(\begin{array}{ccc}
\times & \times &\times\\
\times & 0 &\times \\
\times& \times& 0 
\end{array}\right);
\end{equation}
where $\times$ implies non-zero arbitrary elements of $M_{\nu}$. In the upcoming sections, we study the above-mentioned texture-zero mass matrices in the context of present experimental constraints, including JUNO 2025 as well as future neutrino experiments.

\begin{figure}[h!]
    \centering
    \includegraphics[scale=0.5]{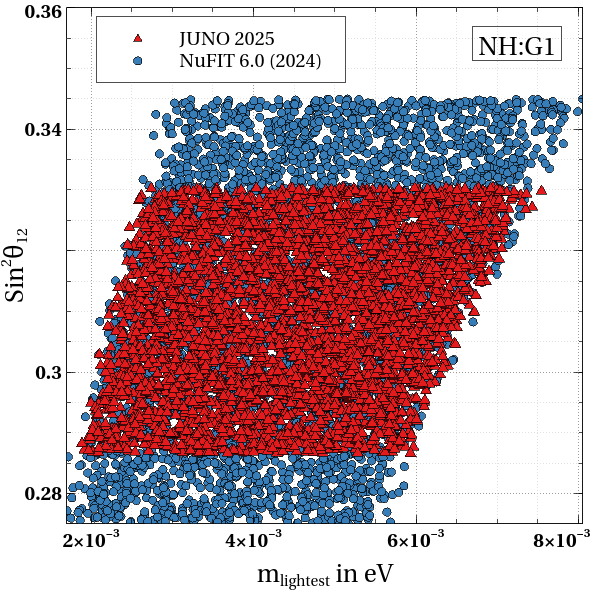}
    \includegraphics[scale=0.5]{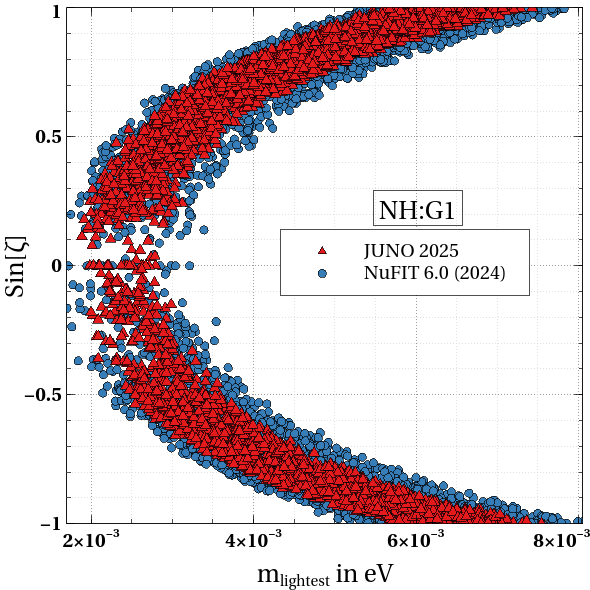}
    \includegraphics[scale=0.5]{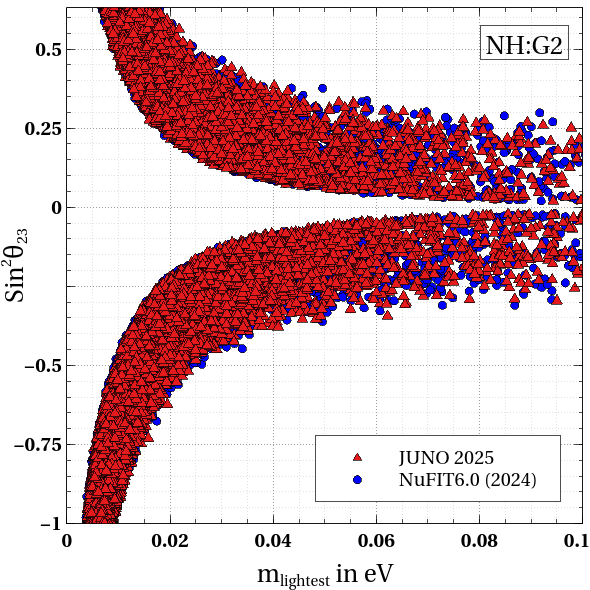}
      \includegraphics[scale=0.5]{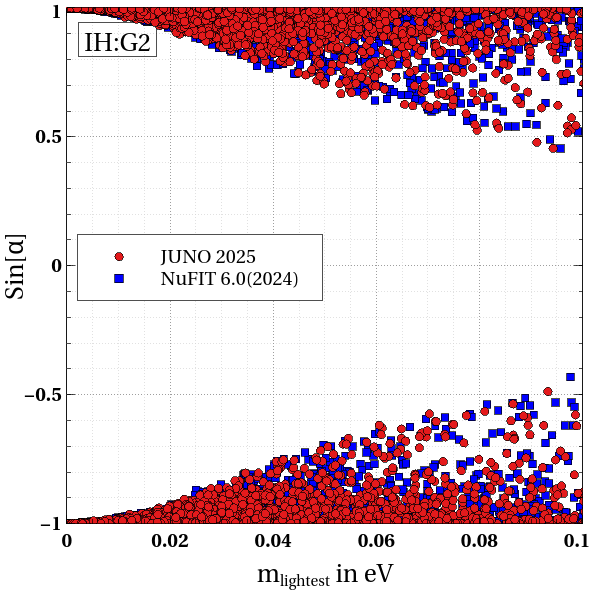}
    \caption{Correlations among neutrino parameters for a few one-zero textures, also showing the impact of JUNO 2025 data over the {\tt NuFit 6.0} (2024) data.}
    \label{fig1}
\end{figure}

\begin{figure}[h!]
    \centering
    \includegraphics[scale=0.45]{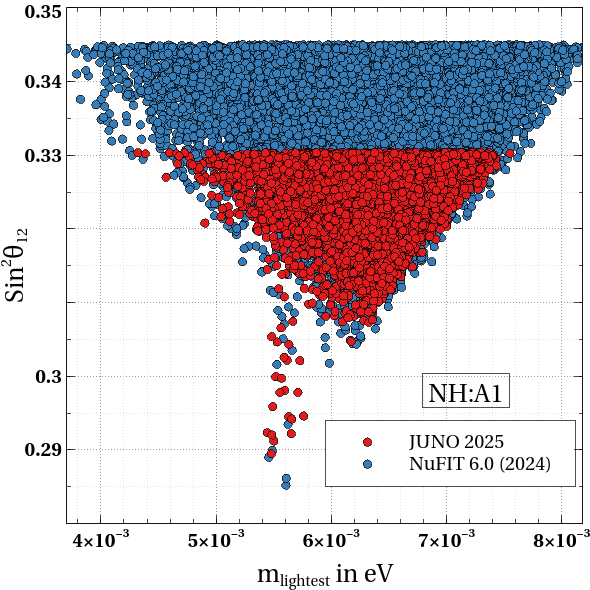}
    \includegraphics[scale=0.45]{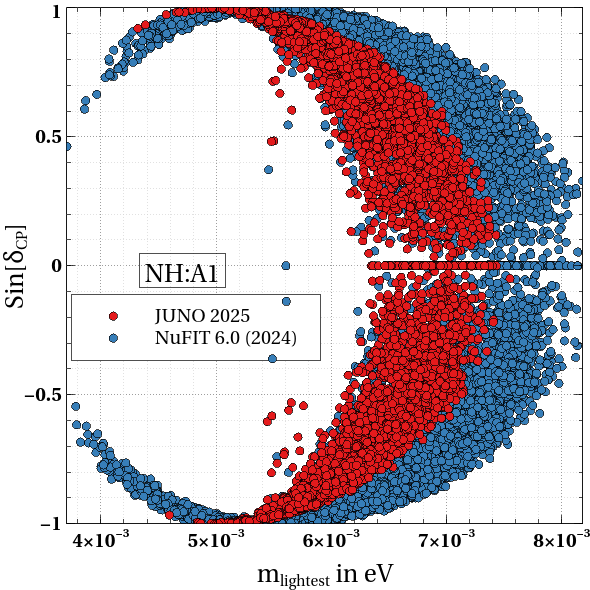}
    \includegraphics[scale=0.45]{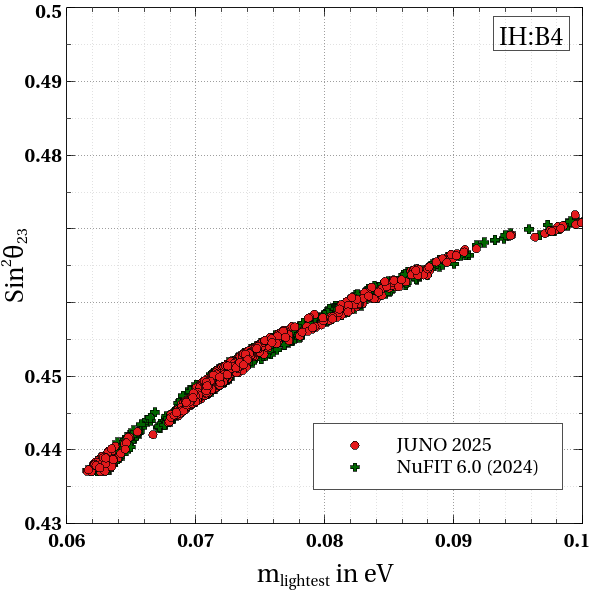}
        \includegraphics[scale=0.45]{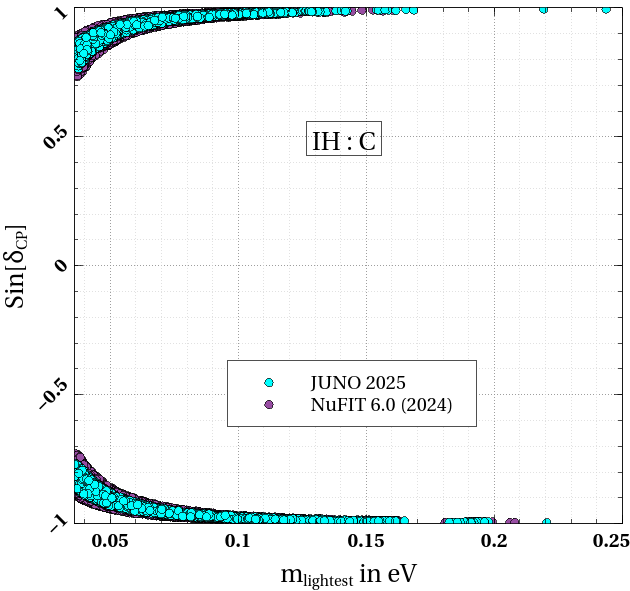}
    \caption{Correlations among neutrino parameters for a few two-zero textures, also showing the impact of JUNO 2025 data over the {\tt NuFit 6.0} (2024) data.}
    \label{fig2}
\end{figure}

\section{Allowed texture-zeros after JUNO's first results}
\label{sec3}
In the diagonal charged lepton basis, $U_{\text{PMNS}}$ is same as the diagonalizing matrix $U_{\nu}$ which can be parametrized as \cite{ParticleDataGroup:2024cfk}
\begin{equation}
U_{\text{PMNS}}=\left(\begin{array}{ccc}
c_{12}c_{13}& s_{12}c_{13}& s_{13}e^{-i\delta_{\text{cp}}}\\
-s_{12}c_{23}-c_{12}s_{23}s_{13}e^{i\delta_{\text{cp}}}& c_{12}c_{23}-s_{12}s_{23}s_{13}e^{i\delta_{\text{cp}}} & s_{23}c_{13} \\
s_{12}s_{23}-c_{12}c_{23}s_{13}e^{i\delta_{\text{cp}}} & -c_{12}s_{23}-s_{12}c_{23}s_{13}e^{i\delta_{\text{cp}}}& c_{23}c_{13}
\end{array}\right) \text{diag}(1, e^{i\alpha}, e^{i(\zeta+\delta_{\text{cp}} )})
\label{matrixPMNS}
\end{equation}
where $c_{ij} = \cos{\theta_{ij}}, \; s_{ij} = \sin{\theta_{ij}}$. $\delta_{\text{cp}}$ is the Dirac CP phase and  $\alpha, \zeta$ are the Majorana CP phases. The presence of Majorana phases plays a crucial role in determining the predictivity of texture-zeros in our neutrino mass matrices. Using the parametric form of the PMNS matrix shown in Eq. \eqref{matrixPMNS}, the Majorana neutrino mass matrix $M_{\nu}$ can be found as
\begin{equation}
 M_{\nu}=U_{\text{PMNS}} M^{\text{diag}}_{\nu}U^T_{\text{PMNS}}
 \label{numatrix}
\end{equation}
where
\begin{equation}
M^{\text{diag}}_{\nu}=\left(\begin{array}{ccc}
m_1& 0&0\\
0& m_2& 0  \\
0& 0 &m_3
\end{array}\right),
\end{equation}
with $m_1, m_2$ and $m_3$ being the three neutrino mass eigenvalues. For the case of a normal hierarchy, the three neutrino mass eigenvalues can be written as 
$$M^{\text{diag}}_{\nu}
= \text{diag}(m_1, \sqrt{m^2_1+\Delta m_{21}^2}, \sqrt{m_1^2+\Delta m_{31}^2})$$ while for the case of inverted hierarchy, they can be written as 
$$M^{\text{diag}}_{\nu} = \text{diag}(\sqrt{m_3^2+\Delta m_{23}^2-\Delta m_{21}^2}, \sqrt{m_3^2+\Delta m_{23}^2}, m_3)$$ 

After parametrizing the neutrino mass matrix elements in terms of these parameters, we numerically solve the texture-zero equations for an appropriate number of parameters while varying the other parameters randomly in the $3\sigma$ range of the latest neutrino global fit data \cite{Esteban:2024eli} as well as JUNO 2025 data \cite{JUNO:2025gmd}. Essentially, the condition $(M_{\nu})_{ab} = 0$ imposes a complex constraint equation that gives rise to correlations between the mixing angles, phases, and mass eigenvalues.

\begin{figure}
    \centering
    \includegraphics[scale=0.5]{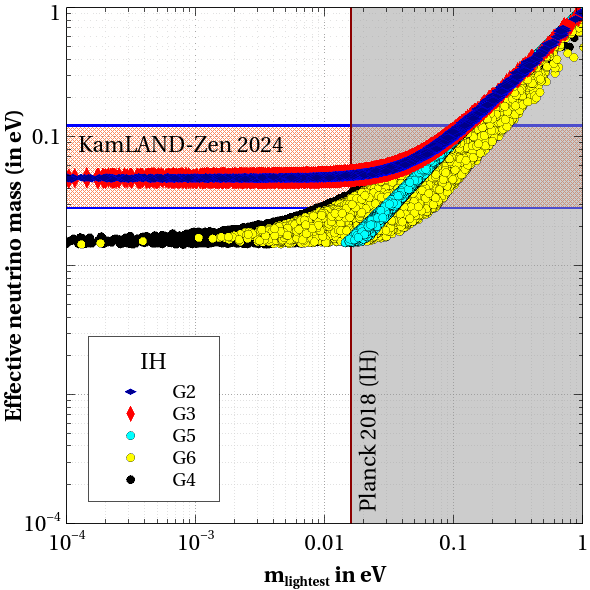}
    \includegraphics[scale=0.5]{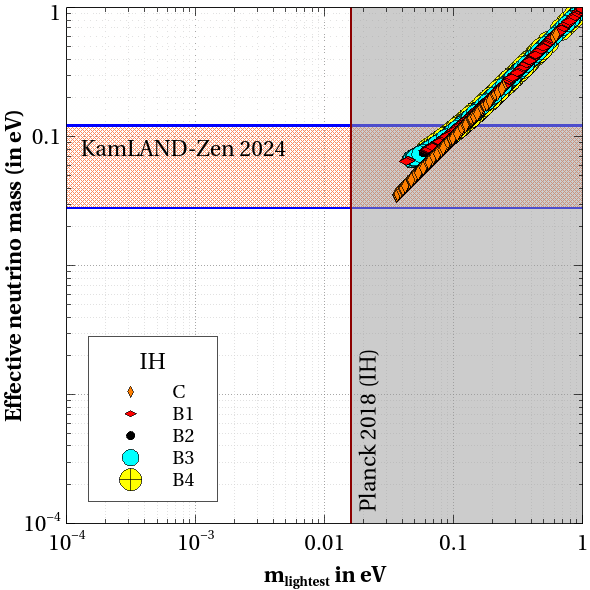}
    \includegraphics[scale=0.5]{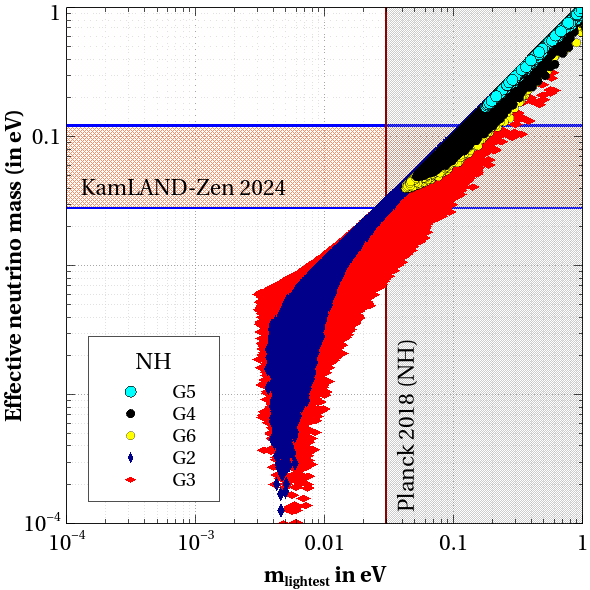}
    \includegraphics[scale=0.5]{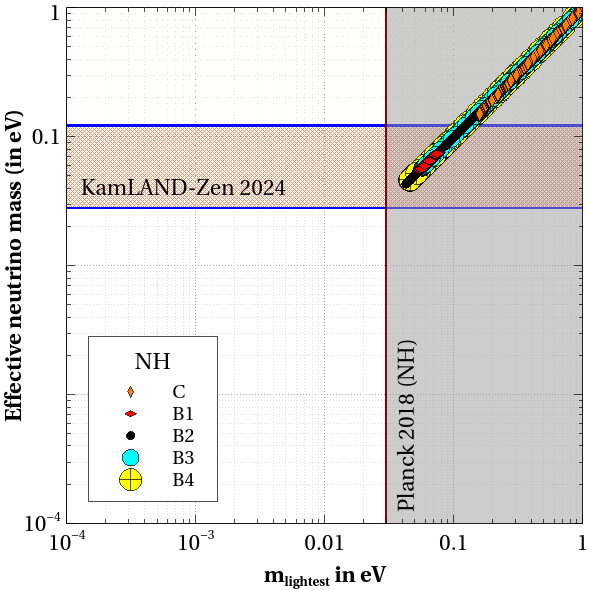}
    \caption{Effective neutrino mass for $0\nu \beta \beta$ as a function of lightest neutrino mass for one-zero and two-zero textures. The orange horizontal patch is the recent KamLAND-Zen upper bound on effective neutrino mass in the range $28-122$ meV \cite{KamLAND-Zen:2024eml} while the vertical grey shaded region is the excluded cosmological bound on the lightest neutrino mass given by PLANCK 2018 data \cite{Planck:2018vyg}.}
    \label{fig:meff}
\end{figure}

Fig. \ref{fig1} and Fig. \ref{fig2} show some of the correlations among different neutrino parameters for a few one-zero and two-zero textures, respectively. Apart from showing the predictive nature of texture-zeros giving interesting correlations, the plots also show the impact of JUNO 2025 data in comparison to the earlier global fit data from {\tt NuFit 6.0}. We can see significantly constrained regions in the allowed parameter space, especially among the lightest neutrino mass and the mixing angles \footnote{During our numerical evaluation, we do not put any upper bound on either the sum of light neutrino masses or the lightest neutrino mass. We allow them to vary and comment on their consistency with PLANCK upper bound in the last columns of Table \ref{tablepart1} and Table \ref{tablepart2}. The lightest neutrino mass bound from cosmology  are $m_1\le0.03 \text{ eV (NH)}$ and $m_3\le0.0157 \text{ eV (IH)}$. }. Fig. \ref{fig:meff} shows the variation of effective neutrino mass for $0\nu \beta \beta$ with the lightest neutrino mass for all the textures. The horizontal band corresponds to KamLAND-Zen upper limit \cite{KamLAND-Zen:2024eml} with the range $28-122$ meV, indicating the uncertainties in nuclear matrix elements. The vertical shaded grey region is disfavored from the cosmological upper bound on neutrino mass \cite{Planck:2018vyg}. Tables \ref{tablepart1} and \ref{tablepart2} list the summary of our results for one-zero and two-zero textures, respectively. We show the validity of these textures with respect to neutrino global fit data, JUNO 2025 data, $0\nu \beta \beta$ bound and cosmological upper limit on sum of absolute neutrino masses $\sum_i \lvert m_i \rvert < 0.12$ eV. While JUNO 2025 data do not change the validity of one-zero textures compared to {\tt NuFit 6.0} data, they do so for two-zero textures. Interestingly, the two-zero texture $B_1$ with NH gets ruled out by JUNO 2025 data, while it is still allowed by {\tt NuFit 6.0} data. Apart from changing the validity of this texture, JUNO 2025 data also changes the allowed neutrino parameter space, as indicated in Fig. \ref{fig1} and Fig. \ref{fig2}, which can be probed in future runs of JUNO or future experiments. 
Below, we consider all the textures allowed by JUNO 2025 neutrino data and bounds from $0\nu \beta \beta$ experiments to study their prospects at DUNE. The details of the texture ruled out by JUNO 2025 data is given in Appendix \ref{appen1}.

\begin{table}
\begin{tabular}{|c|c|c|c|c|}
 \hline
Patterns     & {\tt NuFit 6.0} IH (NH) & JUNO 2025 IH (NH)& $0\nu \beta \beta$ Bound IH (NH)& PLANCK 2018 IH (NH)\\
        \hline \hline
\mbox{$G_1$}     & $\times$($\checkmark$) & $\times$($\checkmark$)  & $\checkmark$($\checkmark$)        &$\times$($\checkmark$) \\
\mbox{$G_2$}     & $\checkmark$($\checkmark$) & $\checkmark$($\checkmark$)  & $\checkmark$($\checkmark$)        & $\checkmark$($\checkmark$)\\ 
\mbox{$G_3$}     & $\checkmark$($\checkmark$)  & $\checkmark$($\checkmark$)     & $\checkmark$($\checkmark$)        &$\checkmark$($\checkmark$)\\
\mbox{$G_4$}     & $\checkmark$($\checkmark$)  & $\checkmark$($\checkmark$) & $\checkmark$($\checkmark$)    &$\checkmark$($\times$)\\
\mbox{$G_5$}     & $\checkmark$($\checkmark$)  & $\checkmark$($\checkmark$)         & $\checkmark$($\times$)    &$\checkmark$($\times$)\\
\mbox{$G_6$}     & $\checkmark$($\checkmark$)  & $\checkmark$($\checkmark$)    & $\checkmark$($\checkmark$)    &$\checkmark$($\times$)\\
        \hline
\end{tabular}
 
\caption{Summary of results for one-zero texture with inverted and normal hierarchy. The symbol $\checkmark$ ($\times$) is used when the particular texture zero mass matrix is (not) consistent with the respective experimental bound.}
\label{tablepart1}
\end{table}

\begin{table}
\begin{tabular}{|c|c|c|c|c|}
 \hline
Patterns     & {\tt NuFit 6.0} IH (NH) & JUNO 2025 IH (NH) & $0\nu \beta \beta$ Bound IH (NH)& PLANCK Bound IH (NH)\\
        \hline \hline
\mbox{$A_1$}     & $\times$($\checkmark$) & $\times$($\checkmark$)  & $\checkmark$($\checkmark$)         &$\times$($\checkmark$) \\
\mbox{$A_2$}     & $\times$($\times$)  & $\times$($\times$) & $\checkmark$($\checkmark$)       & $\times$($\times$)\\ 
\mbox{$B_1$}     & $\checkmark$($\checkmark$) & $\checkmark$($\times$)      & $\checkmark$($\checkmark$)        &$\times$($\times$)\\
\mbox{$B_2$}     & $\checkmark$($\checkmark$)  & $\checkmark$($\checkmark$) & $ \checkmark $($\checkmark$)    &$\times$($\times$)\\
\mbox{$B_3$}     & $\checkmark$($\checkmark$)  & $\checkmark$($\checkmark$)         & $\checkmark$($\checkmark$)    &$\times$($\times$)\\
\mbox{$B_4$}     & $\checkmark$($\checkmark$)  &  $\checkmark$($\checkmark$)   & $\checkmark $($\checkmark$)    &$\times$($\times$)\\
\mbox{$C$}     & $\checkmark$($\checkmark$)  &  $\checkmark$($\checkmark$)   & $\checkmark $($\times$)    &$\times$($\times$)\\
        \hline
\end{tabular}
 
\caption{Summary of results for two-zero texture with inverted and normal hierarchy. The symbol $\checkmark$ ($\times$) is used when the particular texture zero mass matrix is (not) consistent with the respective experimental bound. }
\label{tablepart2}
\end{table}

\section{Implications for long-baseline neutrino experiments}
\label{sec4}
\subsection{DUNE} 
The Deep Underground Neutrino Experiment, or DUNE, is a next-generation facility designed to deliver precision neutrino measurements and probe physics beyond the Standard Model. Utilizing high-intensity proton beams, a sophisticated near-detector system, and a massive far-detector located deep underground, DUNE provides unprecedented sensitivity to oscillation phenomena. It will employ a high-intensity broadband neutrino beam from the LBNF at Fermilab, initially operating at 1.2~MW \cite{DUNE:2015lol}, with simulations detailed in \cite{DUNE:2021cuw}. The beam peaks around $E_\nu \sim 2.5$~GeV, covering the first oscillation maximum. DUNE aims to determine the neutrino mass ordering, discover leptonic CP violation, and precisely measure the oscillation parameters \cite{DUNE:2015lol, DUNE:2020ypp}. Its long baseline enhances sensitivity through matter effects, while also probing proton decay, supernova neutrinos, and physics beyond the Standard Model \cite{DUNE:2020fgq}.

\subsection{T2HK}
The T2HK~\cite{Hyper-KamiokandeProto-:2015xww, Hyper-Kamiokande:2018ofw} is a proposed upgrade of the ongoing T2K experiment in Japan, with neutrinos produced at Tokai using an enhanced J-PARC beam. While the present beam power is about 470~kW, it is expected to exceed 1.3~MW by the time T2HK becomes operational. The upgraded detector, Hyper-Kamiokande (HK), will consist of two identical upright cylindrical tanks, each with a fiducial mass of 187~kt, yielding a total fiducial mass of 374~kt. Although the final configuration of the near detector has not yet been determined, several options are under consideration, including an upgraded ND280 detector and a smaller water-\v{C}erenkov detector similar in concept to the far detector. The far detector will be an advanced version of the Super-Kamiokande (SK)~\cite{Super-Kamiokande:2002weg}, located 295~km from Tokai in the Kamioka mine and positioned 2.5$^\circ$ off the beam axis. This configuration produces a narrow-band neutrino beam peaked at approximately 0.56~GeV, corresponding to the first oscillation maximum.

\subsection{$\chi^2$ analysis} 
In this section, we investigate the sensitivity of one-zero and two-zero textures in the context of the long-baseline neutrino experiments DUNE and T2HK. 
For the calculation of the expected event spectra for DUNE and T2HK, we use the {\tt GLoBES} (General Long Baseline Experiment Simulator) \cite{Huber:2004ka, HUBER2007439} software package. The experiment-specific configurations for each experiment are implemented in GLoBES using the Abstract Experiment Definition Language (AEDL) files, in which the detector specifications, neutrino and antineutrino fluxes, interaction cross sections, energy resolution, and systematic uncertainties are included as per the details provided by the respective collaborations. For DUNE, these details are provided by the collaboration and are described in Ref. \cite{DUNE:2021cuw, DUNE:2020fgq}. For T2HK, we reproduce the signal and background spectra reported by the Hyper-Kamiokande Collaboration using the collaboration-provided flux files and the detector specifications as per the Technical Design Report ~\cite{Hyper-KamiokandeProto-:2015xww}.

To analyze the sensitivity of the combined setup to these textures, we generate simulated data using the current best-fit values from {\tt NuFit 6.0}, with the exception of the parameters $\delta_{\rm CP}$ and $\theta_{23}$. These two parameters are varied over their $3\sigma$ allowed ranges in the data. While testing a particular texture, we vary the numerically calculated oscillation parameters corresponding to the texture hypotheses in the fit. It is to be noted that the lightest neutrino mass plays an important role in our analysis, and we scan it over its phenomenologically allowed range. For each value, we solve the texture-zero equations and obtain the corresponding neutrino masses, mixing angles, and CP-violating phases consistent with the imposed texture-zero conditions. We then test the resulting parameter sets against the neutrino oscillation data, and only the allowed solutions are used as inputs to the {\tt GLoBES} simulations. The $\chi^2$ is then calculated by marginalising over the relevant oscillation parameters, with the JUNO prior included for $\sin^2\theta_{12}$. 
The $\Delta \chi^2$ used in this work is defined as:

 \begin{equation}
\begin{split}
    \Delta\chi^{2}(p^{\text{true}}) = \min_{p^{\text{test}},\eta} \Bigg[ & 2\sum_{i,j,k}^{} \left\{ N_{ijk}^{\text{test}}(p^{\text{test}};\eta) - N_{ijk}^{\text{true}}(p^{\text{true}}) + N_{ijk}^{\text{true}}(p^{\text{true}}) \ln\frac{N_{ijk}^{\text{true}}(p^{\text{true}})}{N_{ijk}^{\text{test}}(p^{\text{test}};\eta)} \right\} \\
    & + \sum_{l} \frac{(p_{l}^{\text{true}} - p_{l}^{\text{test}})^{2}}{\sigma_{p_{l}}^{2}} + \sum_{m} \frac{\eta_{m}^{2}}{\sigma_{\eta m}^{2}} \Bigg].
\end{split}
\end{equation}
The total $\chi^2$ is composed of three main contributions:
\begin{itemize}
    \item Poisson log-likelihood term: compares the predicted event counts across all detector energy bins, oscillation channels ($\nu_e$ appearance, $\nu_\mu$ disappearance), and beam configurations (neutrino and antineutrino runs).
    \item Prior term: incorporates uncertainties from external measurements.
    \item Systematic pull term: accounts for experimental uncertainties through nuisance parameters, each defined with a $1\sigma$ variation.
\end{itemize}

The expected event counts are denoted as $N^{\text{true/test}}_{ijk}$, where:
\begin{align*}
    i &\rightarrow \text{energy bins of the detector}, \\
    j &\rightarrow \text{oscillation channels ($\nu_e$ appearance, $\nu_\mu$ disappearance)}, \\
    k &\rightarrow \text{beam operation modes (neutrino run, antineutrino run)}.
\end{align*}

Thus, $N^{\text{true/test}}_{ijk}$ represents the expected number of events in energy bin $i$, for channel $j$, and beam mode $k$, under either the true or test hypothesis. It should be noted that in the $\chi^2$ analysis, we have computed the minimised $\chi^2$ by fitting the model predictions. We show the plots in the  $\sin^2\theta_{23}$ - $\delta_{\rm CP}$ true parameter space by fixing the other oscillation parameters at the best fit values given in \cite{Esteban:2024eli}. We have also implemented the JUNO results in our analysis by adding prior on $\sin^2\theta_{12}$ as :
\begin{equation}
    \chi^2_{\rm JUNO}(\sin^2{\theta_{12}})=\Big(\frac{\sin^2{\theta_{12}}^{\rm fit}-\sin^2{\theta_{12}}^{\rm bf}}{0.0087}\Big)^{2}
 \end{equation}
 where, $\sin^2{\theta_{12}}^{\rm fit}$ comes from the models and $\sin^2{\theta_{12}}^{\rm bf}$ is the best fit taken from Eq. \eqref{bfjuno}. Similarly, we also add the prior on $\sin^2{\theta_{13}}$ as per {\tt NuFit 6.0} and show the impact of these priors in this analysis.

The values of the systematic uncertainties used for DUNE and T2HK can be found in Table ~\ref{tab1}.
\begin{table}
\centering
\renewcommand{\arraystretch}{1.2}
\begin{tabular}{lcc}
\toprule
\textbf{Experiment} & \textbf{Channel} & \textbf{Signal / Background} \\
\midrule
\multirow{2}{*}{DUNE} 
& $\nu_e \;(\bar{\nu}_e)$ appearance 
& $2\% \;(2\%) \; / \; 5\% \;(5\%)$ \\
& $\nu_\mu \;(\bar{\nu}_\mu)$ disappearance 
& $5\% \;(5\%) \; / \; 5\% \;(5\%)$ \\
\midrule
\midrule
\multirow{2}{*}{T2HK} 
& $\nu_e \;(\bar{\nu}_e)$ appearance 
& $3.2\% \;(3.9\%) \; / \; 5\% \;(5\%)$ \\
& $\nu_\mu \;(\bar{\nu}_\mu)$ disappearance 
& $3.6\% \;(3.6\%) \; / \; 5\% \;(5\%)$ \\
\bottomrule
\end{tabular}
\caption{Normalization uncertainties ($\sigma_{\eta}$) for signal and background event rates in the oscillation channels considered. Values in parentheses denote antineutrino mode.}
\label{tab1}
\end{table}

\subsection{Analysis of the results}
In this subsection, we discuss the prospects for further constraining the texture zeros at DUNE and T2HK. We address the following questions: how effectively can different one-zero and two-zero textures be constrained, assuming a total data-taking period of 13 years at DUNE and 10 years at T2HK? How does the recent JUNO measurement of $\theta_{12}$ impact the allowed parameter space when incorporated independently, and when combined with reactor constraints? 

Here we present our results in $\sin^{2}\theta_{23}$--$\delta_{\rm CP}$ (True) plane. We have also highlighted the globally allowed regions (blue and red contours) of the parameter space. We have observed that all models predict highly correlated and restricted regions in the $\sin^{2}\theta_{23}$--$\delta_{\rm CP}$ plane, which are mostly consistent with the current $3\sigma$ {\tt NuFit 6.0} (NH) bounds in NH mode. 
\begin{figure}
    \centering
        \includegraphics[width=0.48\linewidth]{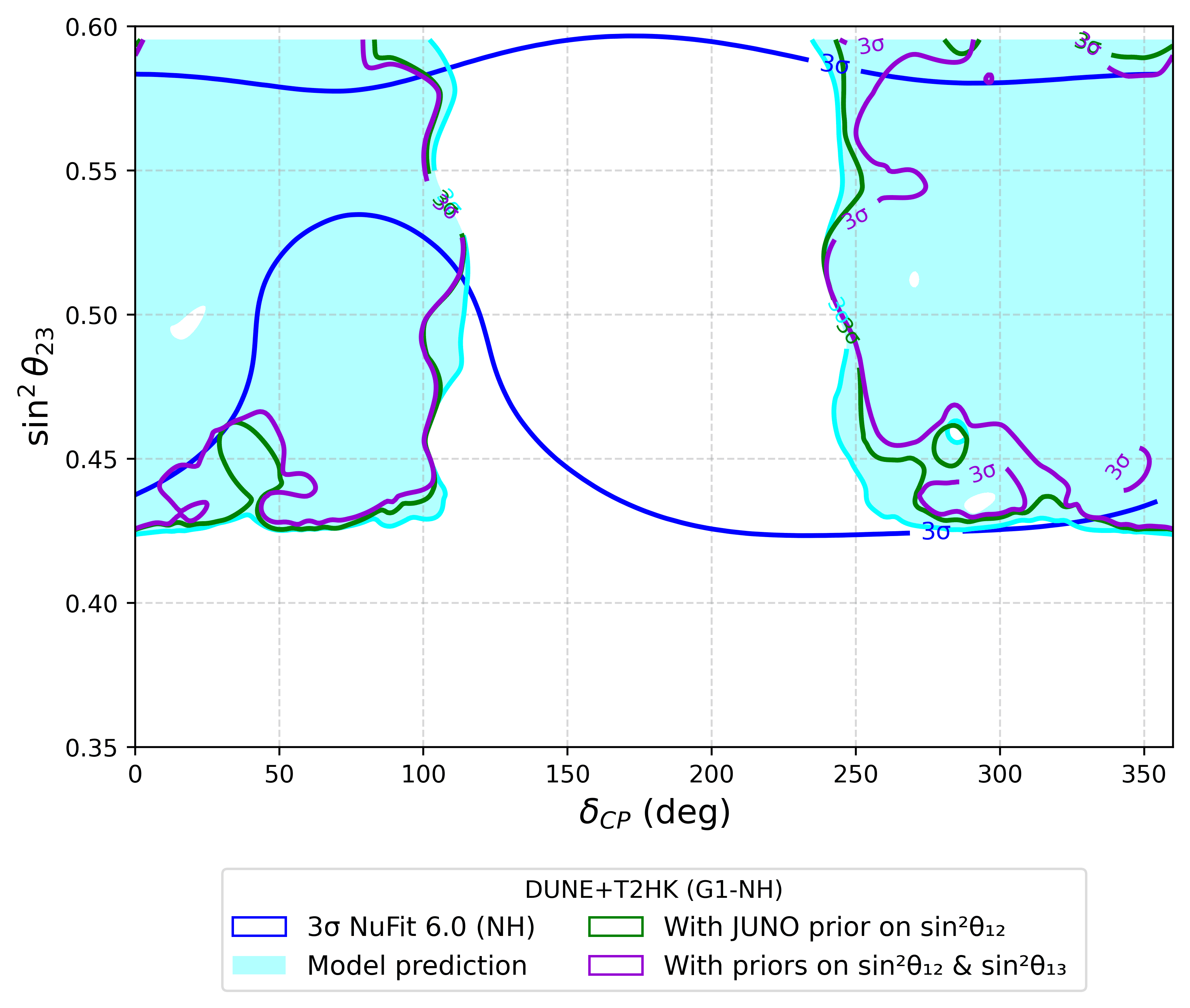}
            \includegraphics[width=0.48\linewidth]{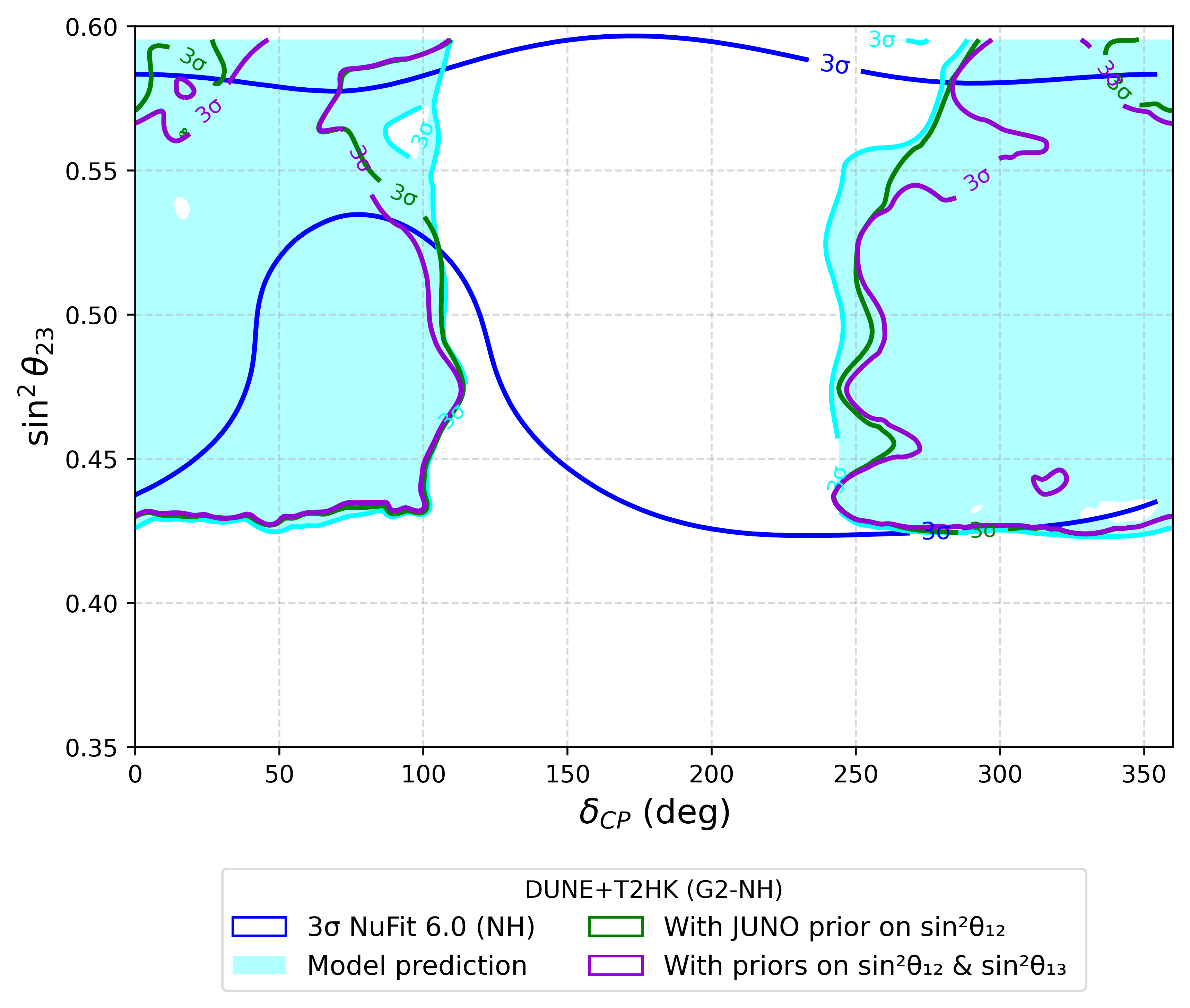}
             \includegraphics[width=0.48\linewidth]{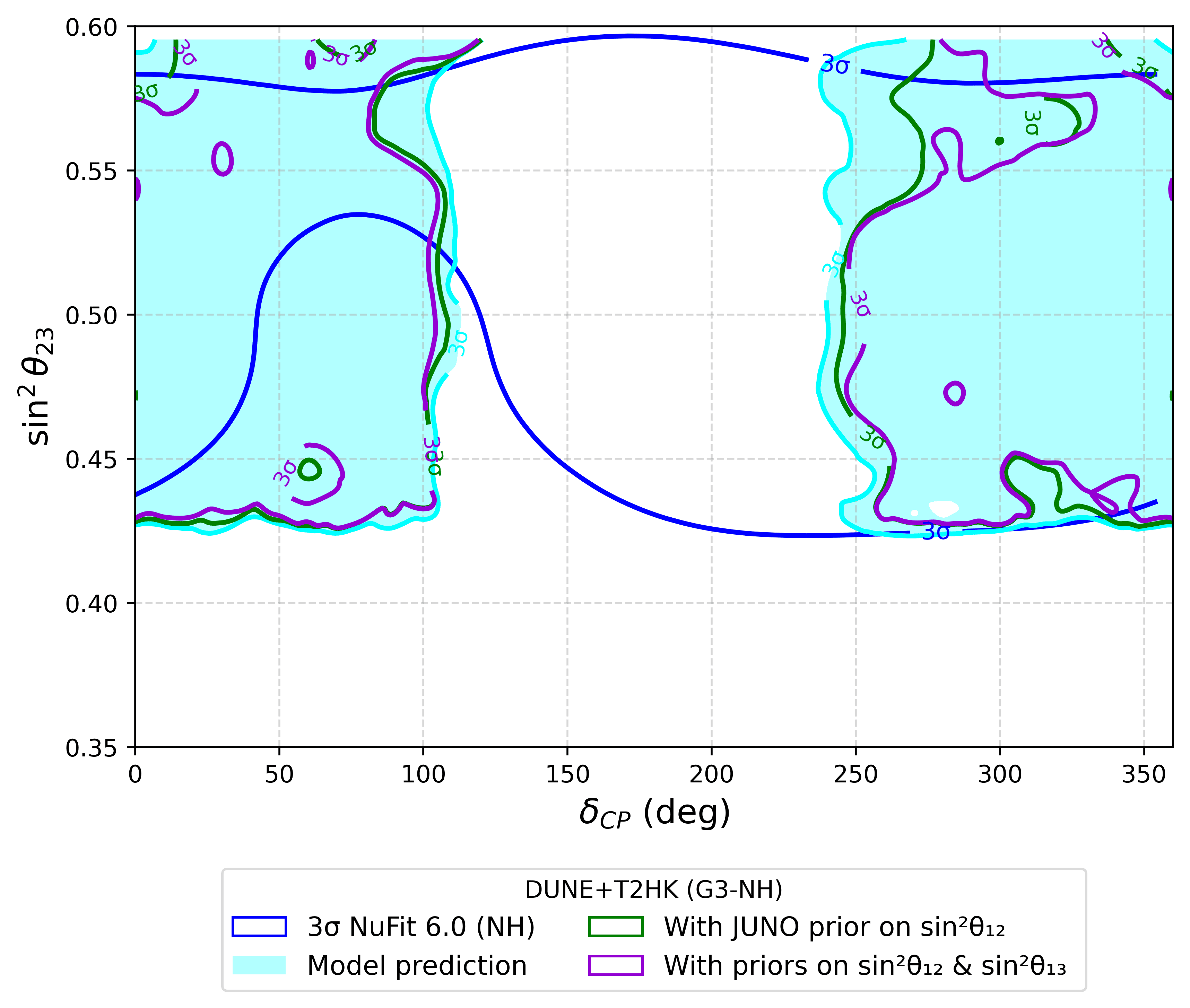}
                 \includegraphics[width=0.48\linewidth]{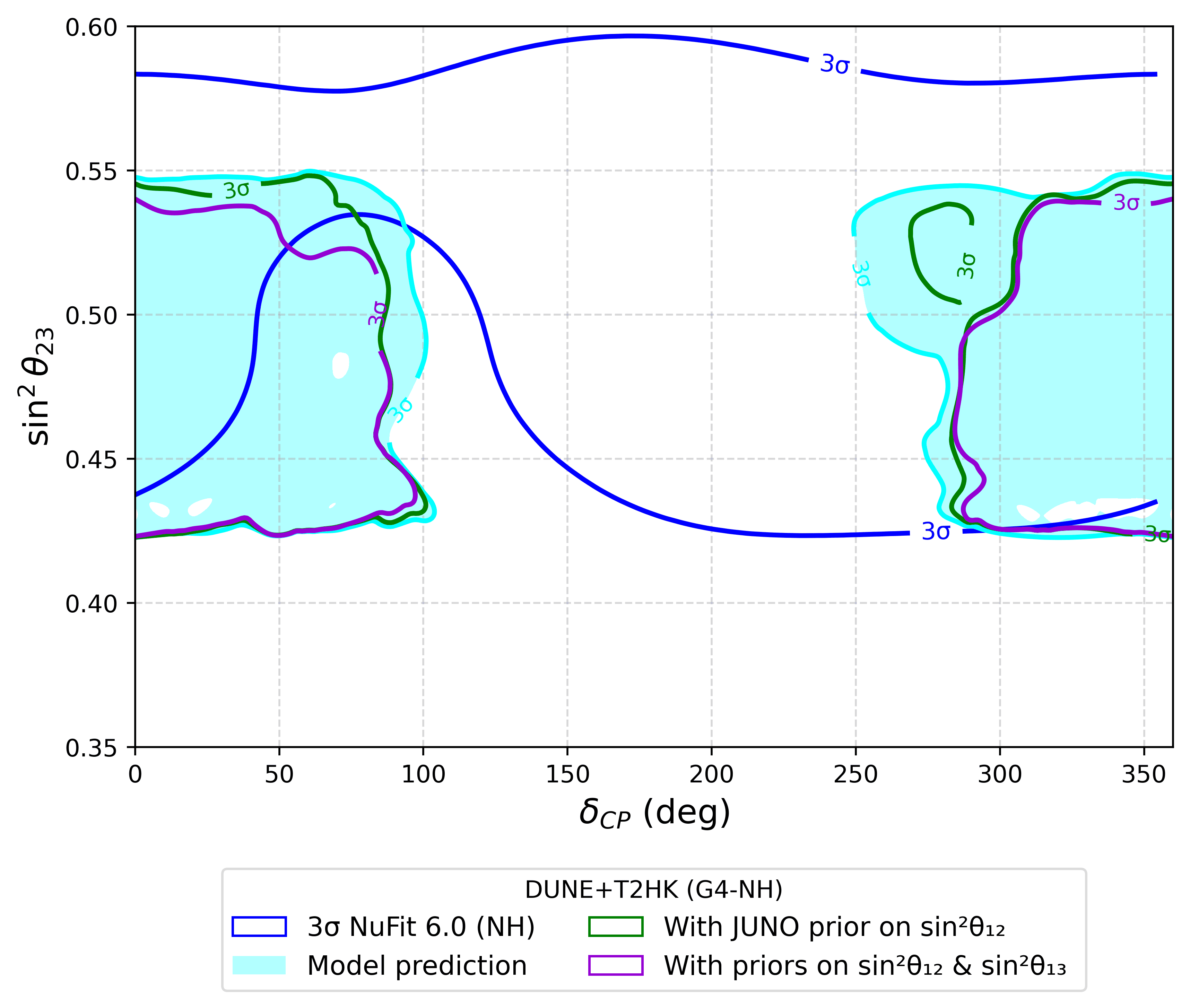}
                 \includegraphics[width=0.48\linewidth]{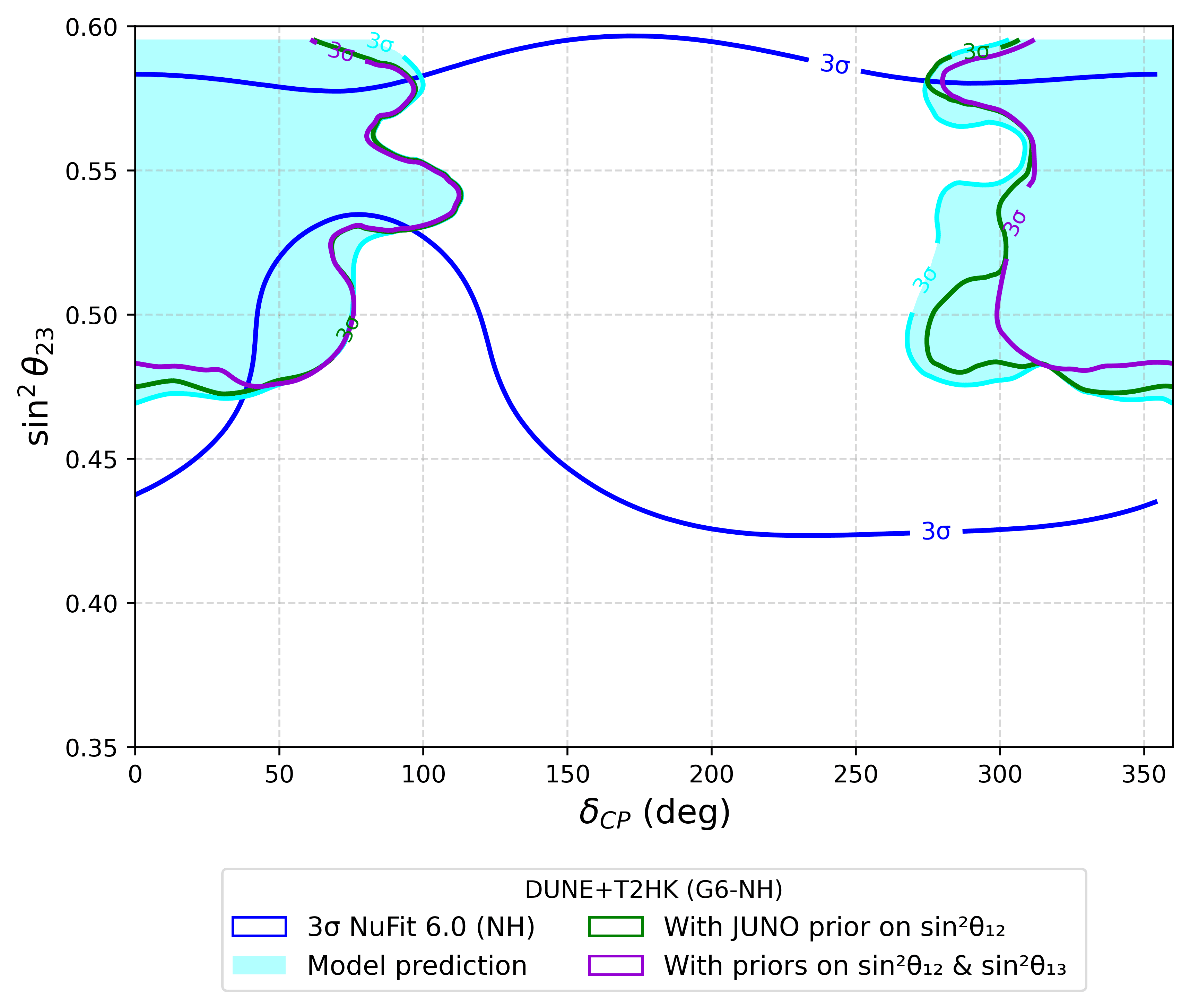}
 \caption{$3\sigma$ allowed regions in the $\sin^{2}\theta_{23}$--$\delta_{\rm CP}$ plane for one-zero textures $G_{1}$--$G_{4}, G_6$ (NH) at DUNE+T2HK. Cyan shaded areas indicate model predictions, while blue contours show the {\tt NuFit 6.0} global-fit constraints. The impact of external information is illustrated by including the JUNO prior on $\sin^{2}\theta_{12}$ (green) and combined priors on $\sin^{2}\theta_{12}$ and $\sin^{2}\theta_{13}$ (purple). }
\label{1zero-NH-comb}
\end{figure}
From Fig.~\ref{1zero-NH-comb}, we observe that compared to the present global-fit constraints, DUNE+T2HK excludes a significantly larger fraction of the model-predicted parameter space, highlighting its enhanced capability to test and discriminate among these scenarios. The allowed regions for the two $G_4$ and $G_6$ textures are significantly more restricted than those of the other textures, indicating that DUNE+T2HK excludes a larger fraction of the parameter space in these two one-zero texture scenarios. The inclusion of the JUNO prior on $\sin^{2}\theta_{12}$ leads to a modest tightening of the allowed regions, while an additional prior on $\sin^{2}\theta_{13}$ produces further improvement in $G_4$ and $G_6$. This demonstrates that the dominant sensitivity arises from $\delta_{\rm CP}$ and $\theta_{23}$, with $\theta_{12}$ playing a subleading role in long-baseline experiments. 

\begin{figure}
    \centering
            \includegraphics[width=0.48\linewidth]{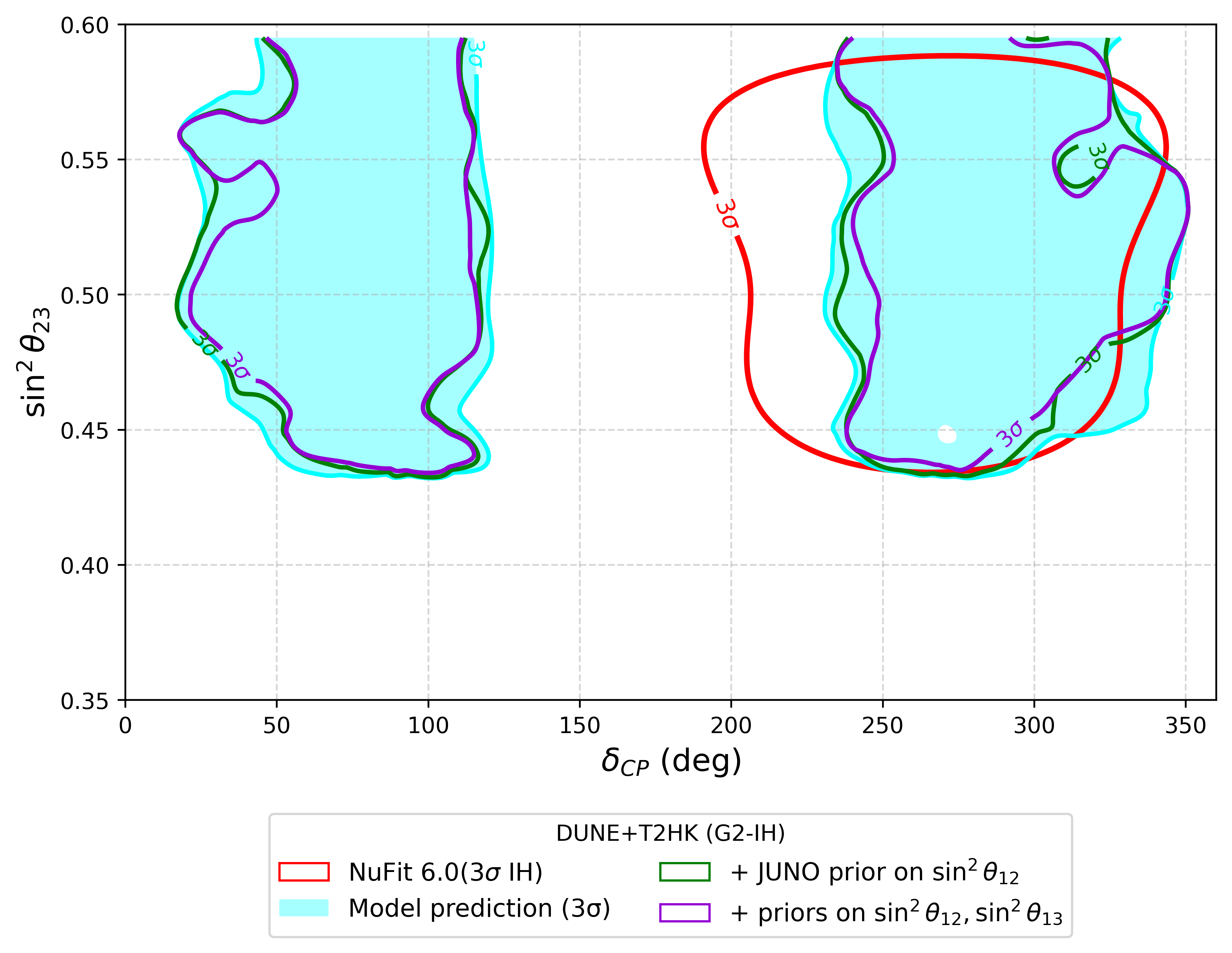}
             \includegraphics[width=0.48\linewidth]{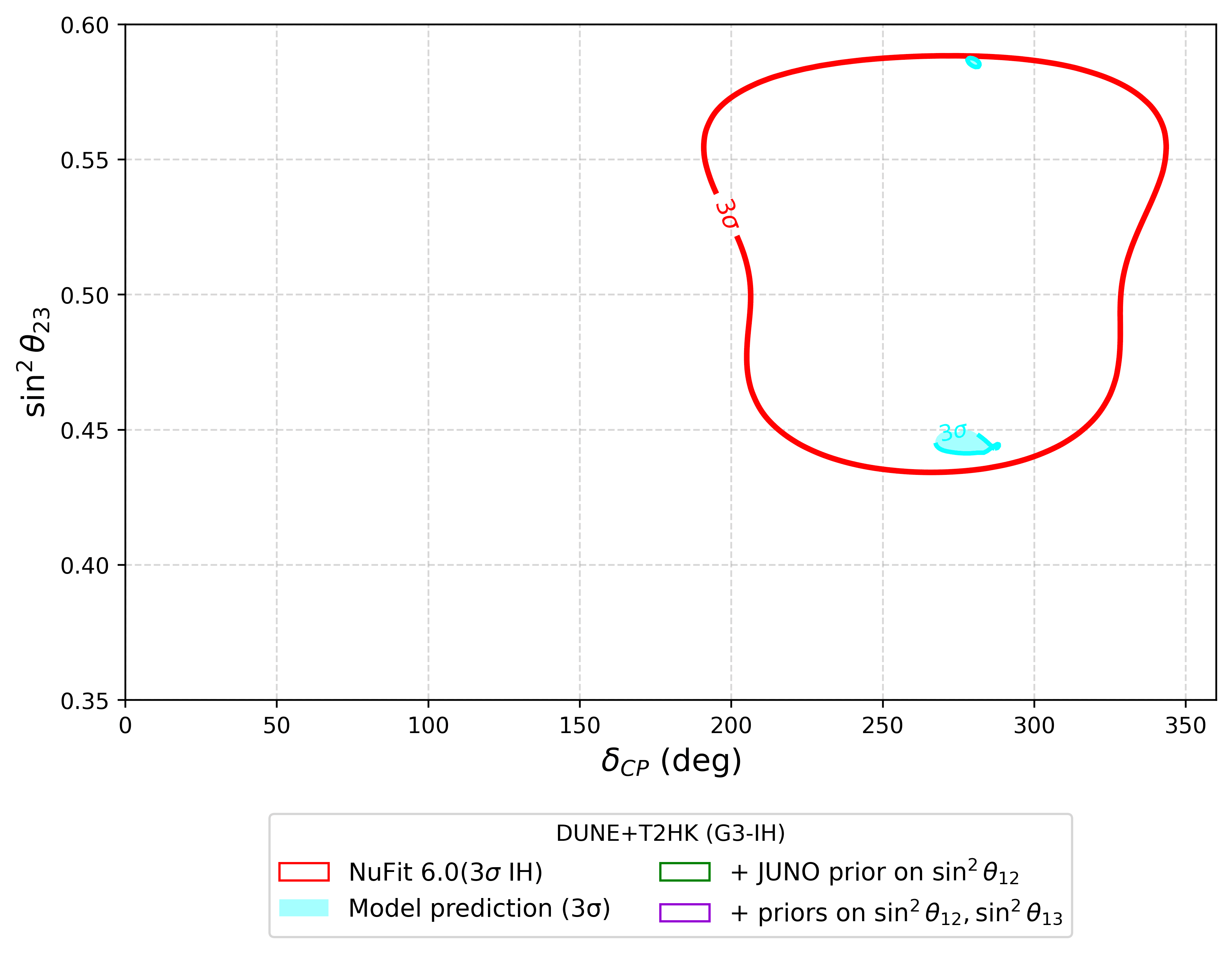}
                 \includegraphics[width=0.48\linewidth]{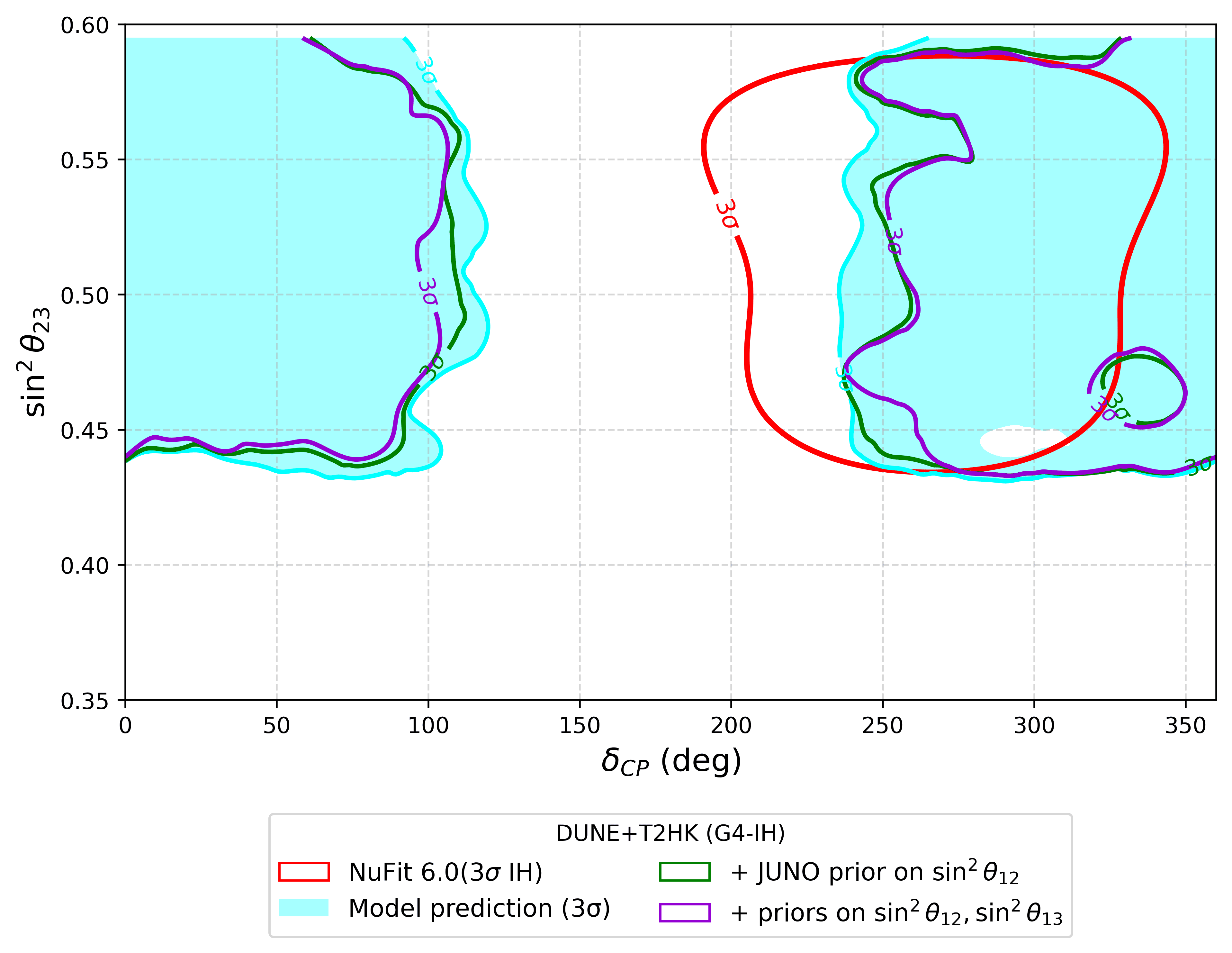}
                \includegraphics[width=0.48\linewidth]{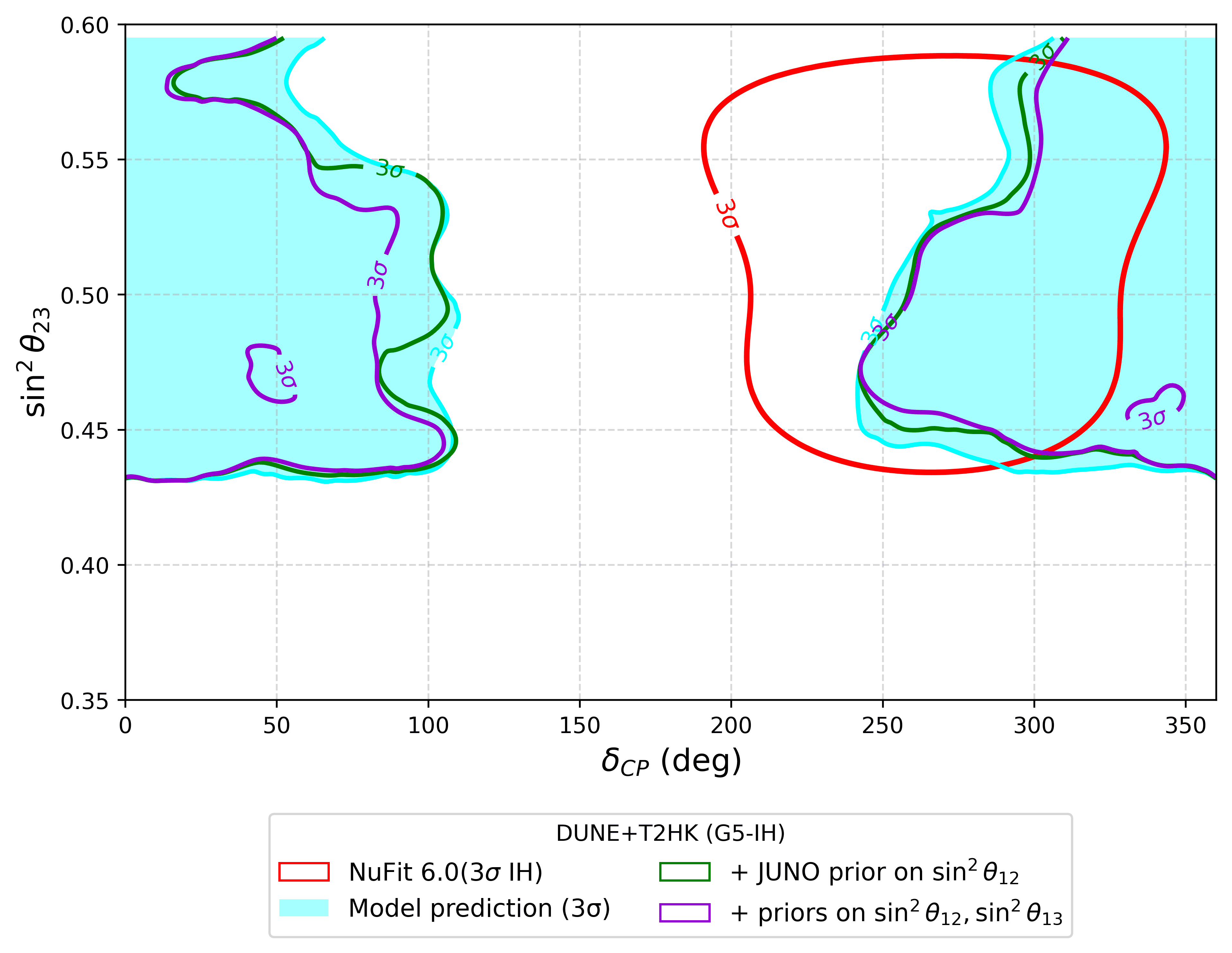}
                 \includegraphics[width=0.48\linewidth]{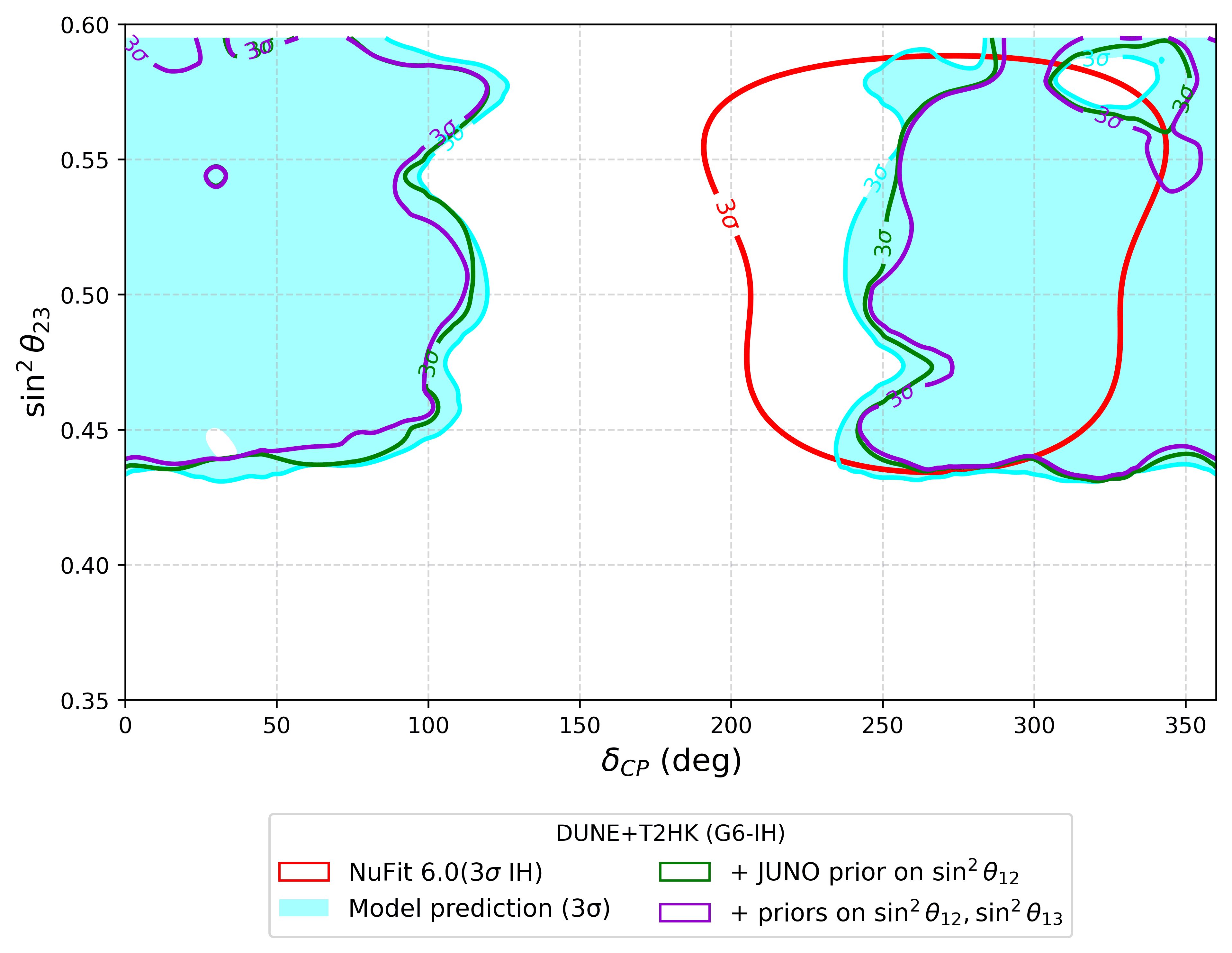}
 \caption{$3\sigma$ allowed regions in the $\sin^{2}\theta_{23}$--$\delta_{\rm CP}$ plane for one-zero textures $G_{2}$--$G_{6}$ (IH) at DUNE+T2HK.  Cyan-shaded areas indicate model predictions, while red contours show the {\tt NuFit 6.0} global-fit constraints. The impact of JUNO prior on $\sin^{2}\theta_{12}$ (green) and combined priors on $\sin^{2}\theta_{12}$ and $\sin^{2}\theta_{13}$ (purple) is also illustrated. }
\label{1zero-IH-comb}
\end{figure}

In the inverted-hierarchy case as shown in Fig. \ref{1zero-IH-comb}, the current $3\sigma$ {\tt NuFit 6.0} constraints (the red contour) allow only non-degenerate regions in the $\sin^{2}\theta_{23}$--$\delta_{\rm CP}$ plane. In contrast, for each texture in the IH mode, we obtain two bands of allowed regions. For $G_2, G_4, G_5$ and $G_6$, a major portion of one of the allowed bands fits around the 3$\sigma$ globally allowed regions (red contour), while the other band appears in the globally ruled out region of the parameter space. On the other hand, the texture $G_3$ is very tightly constrained. In this case, we observe two small isolated allowed regions within the globally allowed region. Interestingly, a substantial fraction of the globally allowed parameter space is ruled out in this case. This demonstrates that, while most one-zero textures exhibit enhanced coverage of the allowed regions in the inverted hierarchy, $G_3$ stands out as the most restrictive and testable scenario. For most of the one-zero textures, the inclusion of external priors leads to additional tightening around the boundaries of the contours. However, for the $G_3$ texture, the inclusion of JUNO and reactor constraints eliminates the allowed parameter space completely. 
\begin{figure}
    \centering
    \includegraphics[width=0.48\linewidth]{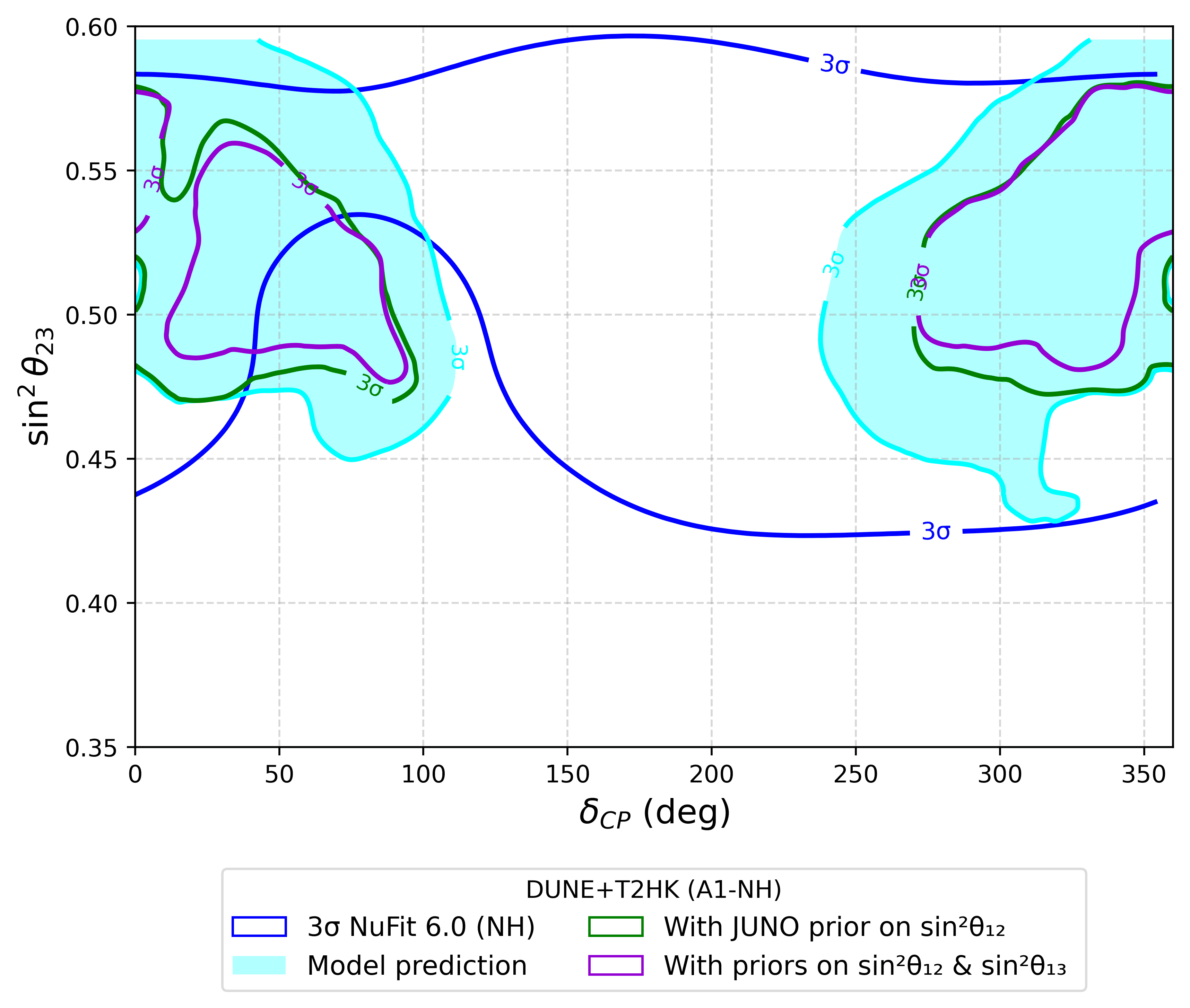}
            \includegraphics[width=0.48\linewidth]{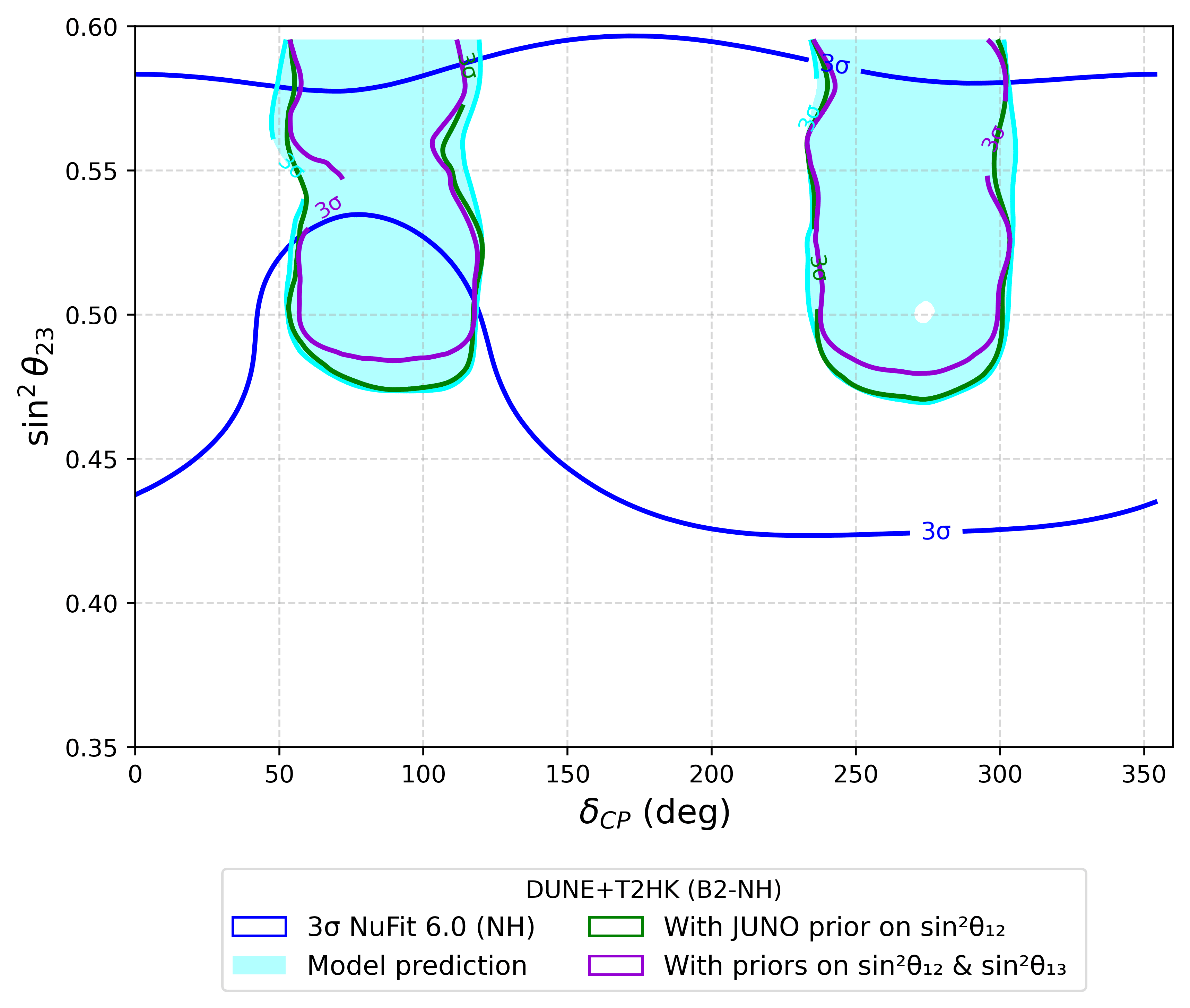}
             \includegraphics[width=0.48\linewidth]{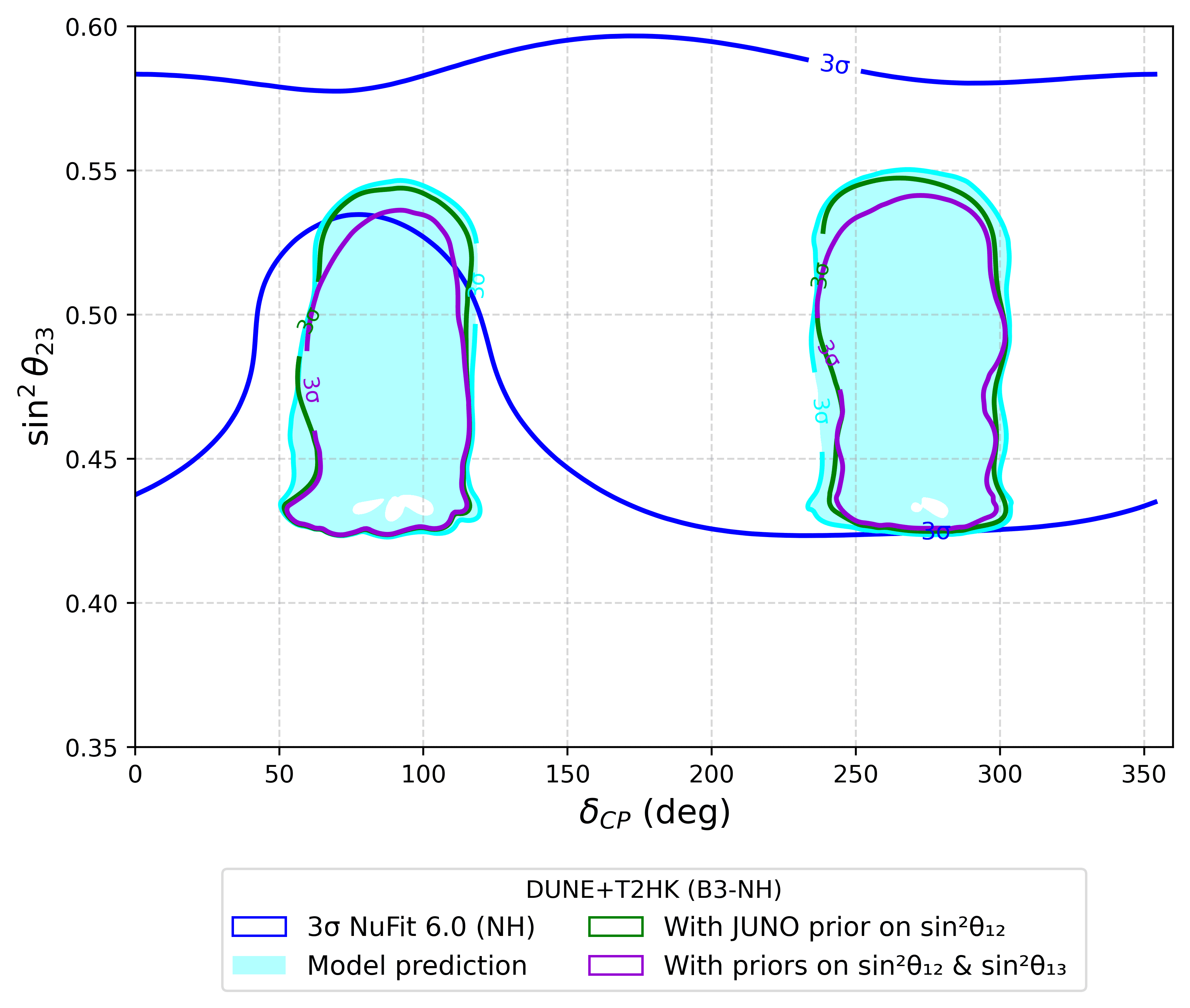}
                 \includegraphics[width=0.48\linewidth]{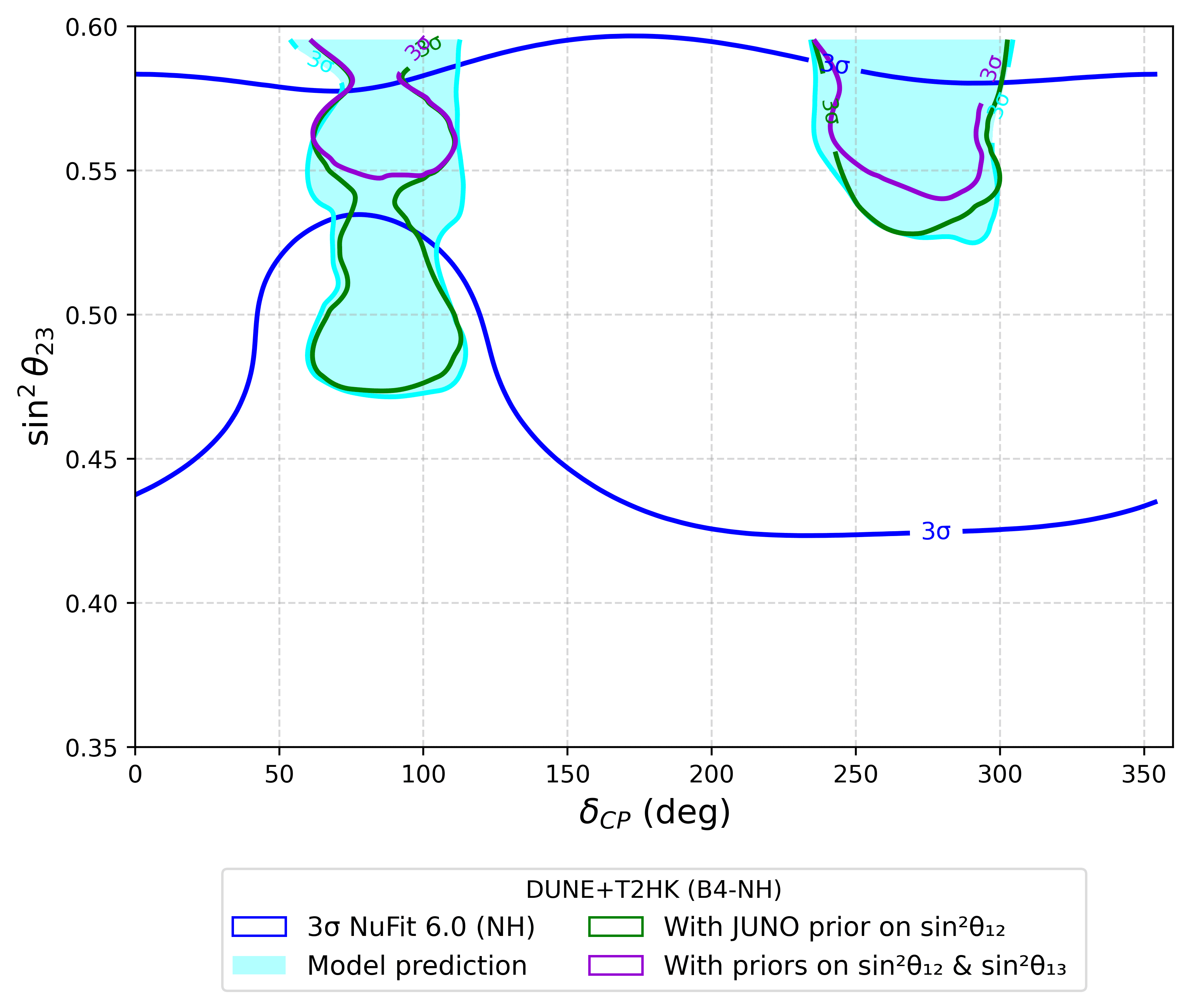}

 \caption{$3\sigma$ allowed regions in the $\sin^{2}\theta_{23}$--$\delta_{\rm CP}$ plane for two-zero textures $A_1$ (NH), $B_2-B_4$ (NH) at DUNE+T2HK.  Color code is the same as in Fig. \ref{1zero-NH-comb}.}
    \label{2zero-NH-comb}
\end{figure}
Fig.~\ref{2zero-NH-comb} shows the allowed regions in the $\sin^{2}\theta_{23}$--$\delta_{\rm CP}$ plane for two-zero textures, namely, $A_1, B_2-B_4$ assuming normal mass hierarchy. In all the cases, the model-predicted regions are very much consistent with the current $3\sigma$ {\tt NuFit 6.0} (NH) bounds, indicating compatibility with present global oscillation data. The predictions typically appear as two correlated allowed bands, reflecting strong texture-induced correlations between $\delta_{\rm CP}$ and $\theta_{23}$. Among the textures considered, $B_2, B_3$ and $B_4$ exhibit more localized allowed regions, while $A_1$ allows a comparatively broader parameter space. The inclusion of the JUNO prior on $\sin^{2}\theta_{12}$ leads to a moderate tightening of the contours, whereas the additional reactor constraint on $\sin^{2}\theta_{13}$ results in a further reduction of the allowed regions. 
\begin{figure}
    \centering
    \includegraphics[width=0.48\linewidth]{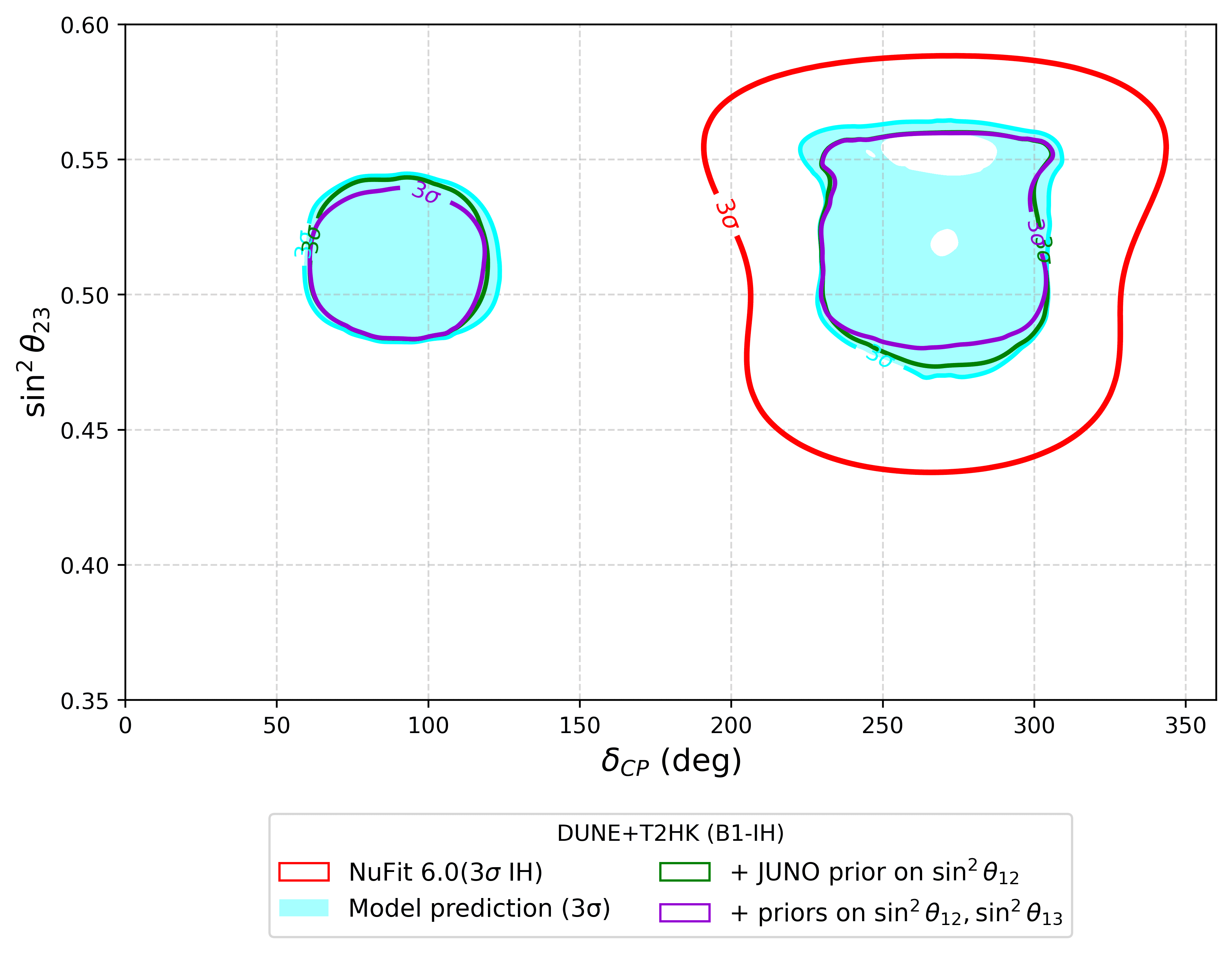}
            \includegraphics[width=0.48\linewidth]{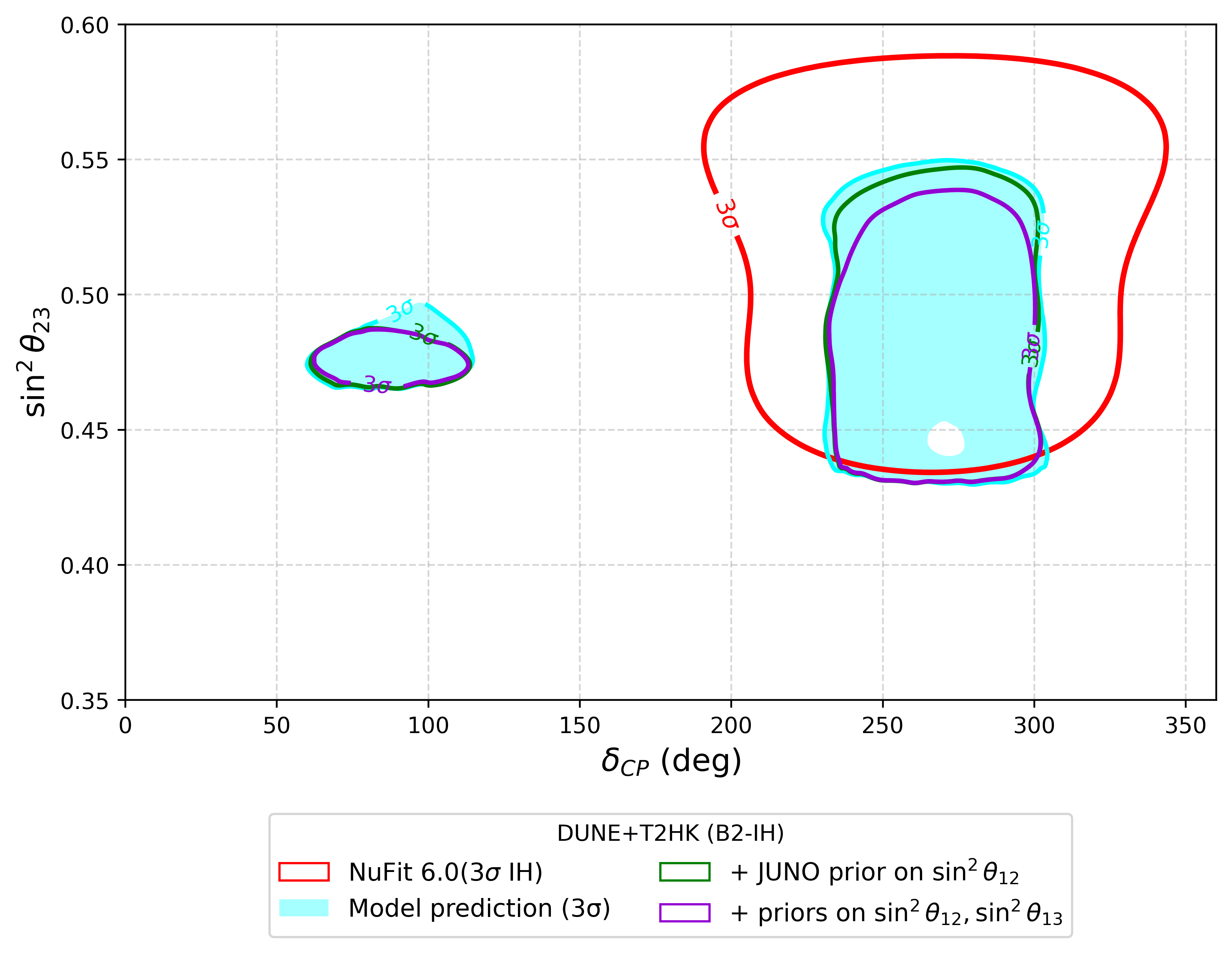}
             \includegraphics[width=0.48\linewidth]{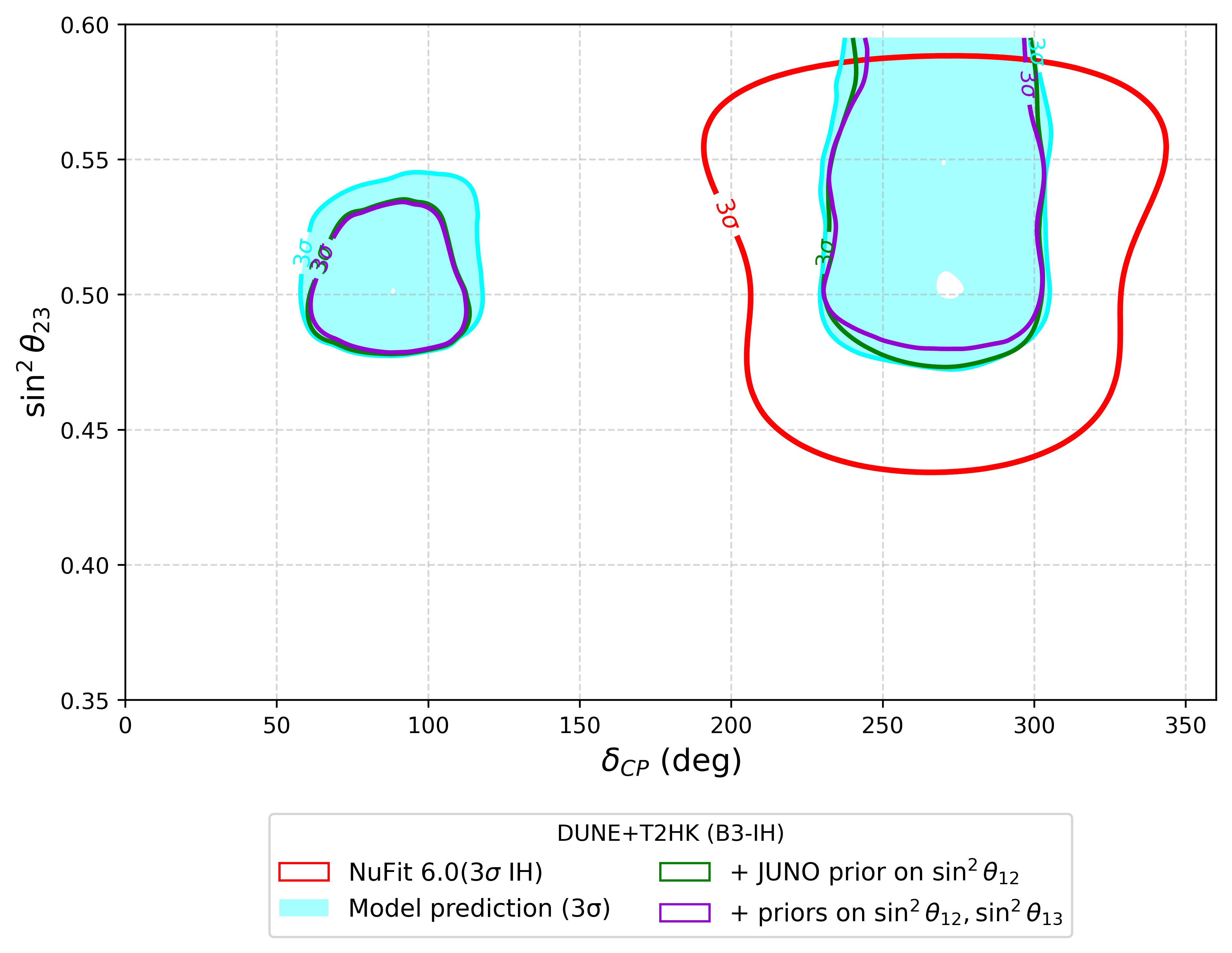}
                 \includegraphics[width=0.48\linewidth]{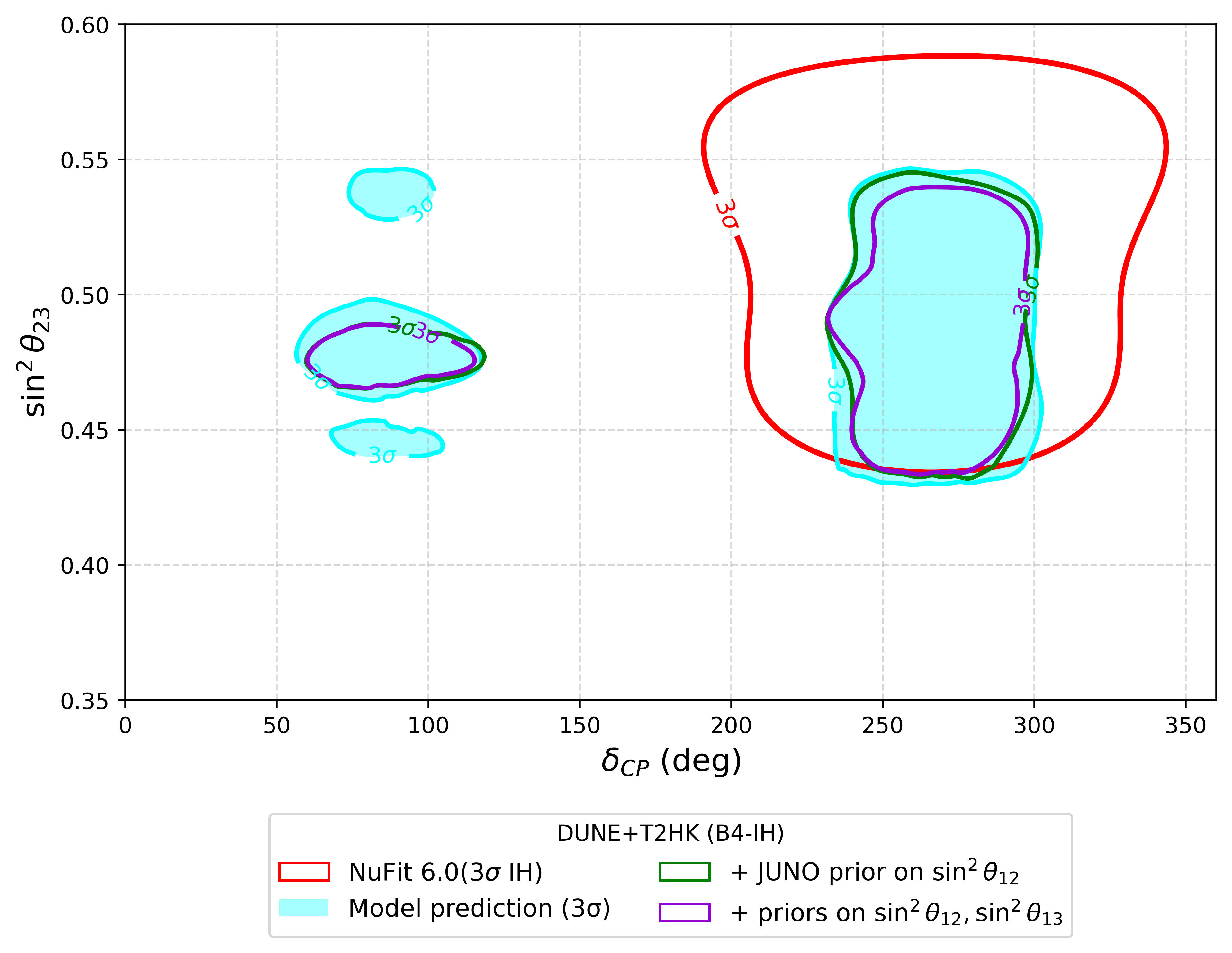}
                   \includegraphics[width=0.48\linewidth]{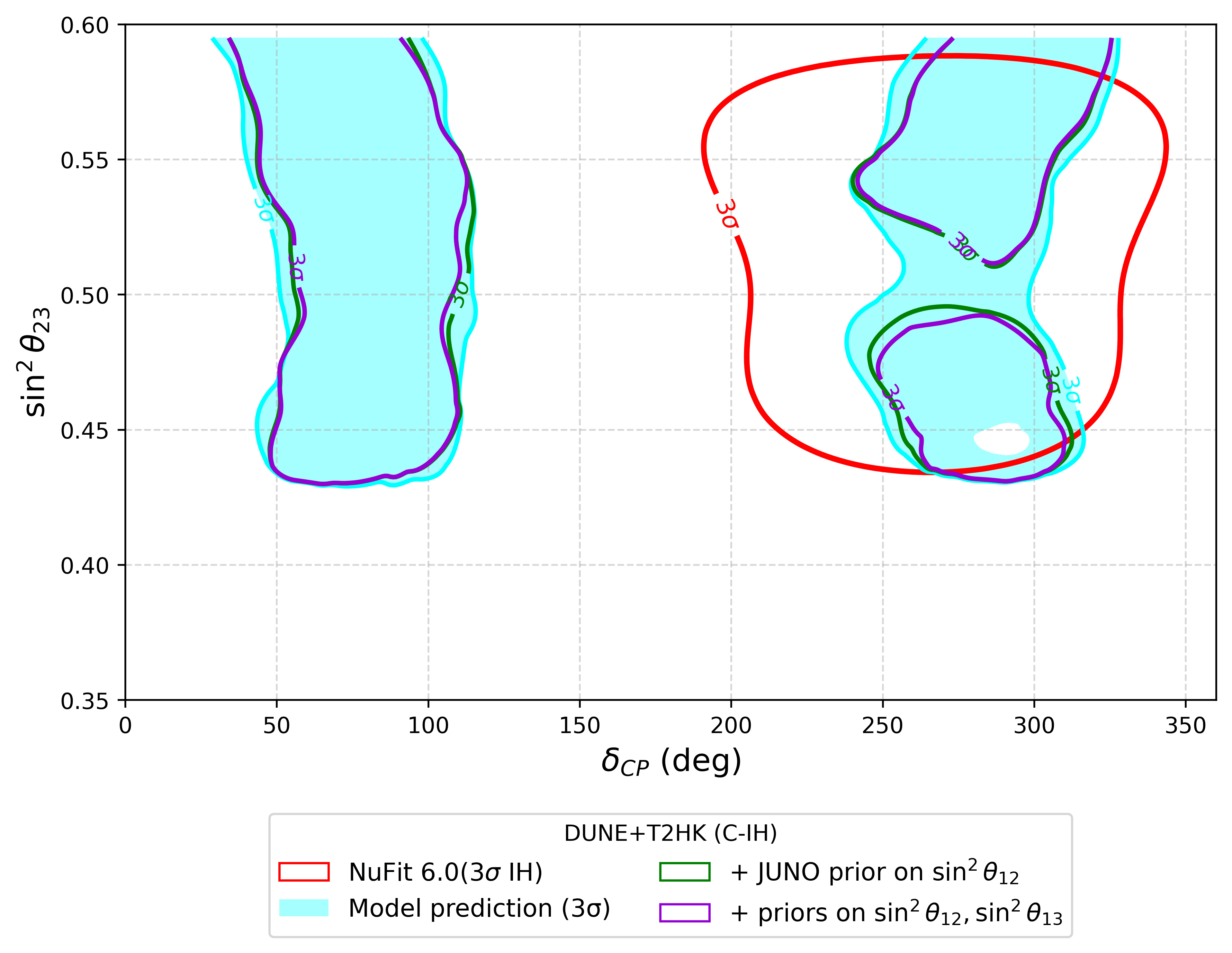}

 \caption{$3\sigma$ allowed regions in the $\sin^{2}\theta_{23}$--$\delta_{\rm CP}$ plane for two-zero textures $B_2-B_4$ (IH) and $C$ (IH) at DUNE+T2HK.  Color code is the same as in Fig. \ref{1zero-IH-comb}.}
    \label{2zero-IH-comb}
\end{figure}

Fig.~\ref{2zero-IH-comb} illustrates the allowed regions in the $\sin^{2}\theta_{23}$--$\delta_{\rm CP}$ plane for the B-type two-zero textures ($B_1-B_4$) in the inverted mass hierarchy. The overall behaviour of the allowed regions is qualitatively similar to that observed for the one-zero textures in the inverted hierarchy; however, in this case, the allowed parameter space is more restricted. The inclusion of the JUNO prior on $\sin^{2}\theta_{12}$ results in a noticeable reduction of the allowed regions, particularly near their boundaries, whereas the additional reactor constraint on $\sin^{2}\theta_{13}$ yields only a mild further effect except $B_4$. As seen in $B_4$ texture, the addition of these priors substantially reduces the size of one of the allowed regions, leading to tighter constraints when JUNO and reactor information are taken into account.

\section{Conclusion}
\label{sec5}
In this work, we have revisited the viability of Majorana neutrino textures in the neutrino mass matrix, assuming the charged lepton mass matrix to be diagonal. Using the latest JUNO 2025 results, we constrain the previously allowed one-zero and two-zero textures further in terms of the allowed parameter space. While the validity of one-zero textures is not affected by JUNO results, we find that one of the two-zero textures currently allowed by global-fit data gets ruled out by JUNO. While PLANCK 2018 bounds on the sum of absolute neutrino masses rule out most of these textures, we adopt a conservative approach with regard to cosmology bounds and study all the textures currently allowed by JUNO data and constraints from neutrinoless double beta decay from the perspective of future long-baseline neutrino experiments like DUNE. We then extend the analysis to the joint DUNE+T2HK configuration to quantify the improvement arising from their combined sensitivity.
More specifically, we have examined the implications of one-zero and two-zero neutrino mass textures in the $\sin^{2}\theta_{23}$--$\delta_{\rm CP}$ plane using the projected sensitivity of DUNE and T2HK, supplemented with external information from JUNO and reactor experiments. We find that most of these textures predict highly correlated and restricted regions that are consistent with current global-fit constraints. This also helps these future experiments to discriminate further among these textures. Future JUNO data will further tighten the restrictions on these textures, and the combined analysis of T2HK and DUNE excludes much of the remaining allowed parameter space of these texture-zero scenarios. While our analysis has been agnostic about UV completions behind these textures, narrowing down the list of allowed textures will also constrain the possible symmetry realisations in a UV-complete setup.

\acknowledgments
The work of D.B. is supported by the Science and Engineering Research Board (SERB), Government of India grants MTR/2022/000575 and CRG/2022/000603. D.D. acknowledges support from the Focus Area Science Technology Summer Fellowship 2025 of the three National Science Academies of India, which enabled a research visit to the Physical Research Laboratory, Ahmedabad.

\appendix
\section{Texture ruled out by JUNO data}
\label{appen1}
Before JUNO, $\theta_{12}$ and $\Delta m_{21}^2$ had larger uncertainties, allowing the two-zero texture to "fit" within the broader allowed parameter space. Recent JUNO results have substantially reduced these uncertainties, drastically shrinking the allowed parameter space. Consequently, upon implementing the JUNO data in our analysis, the previously viable two-zero texture ($B_1$) no longer possesses parameter points consistent with all global oscillation data and is therefore excluded.

\begin{figure}[h!]
    \centering
    \includegraphics[width=0.5\linewidth]{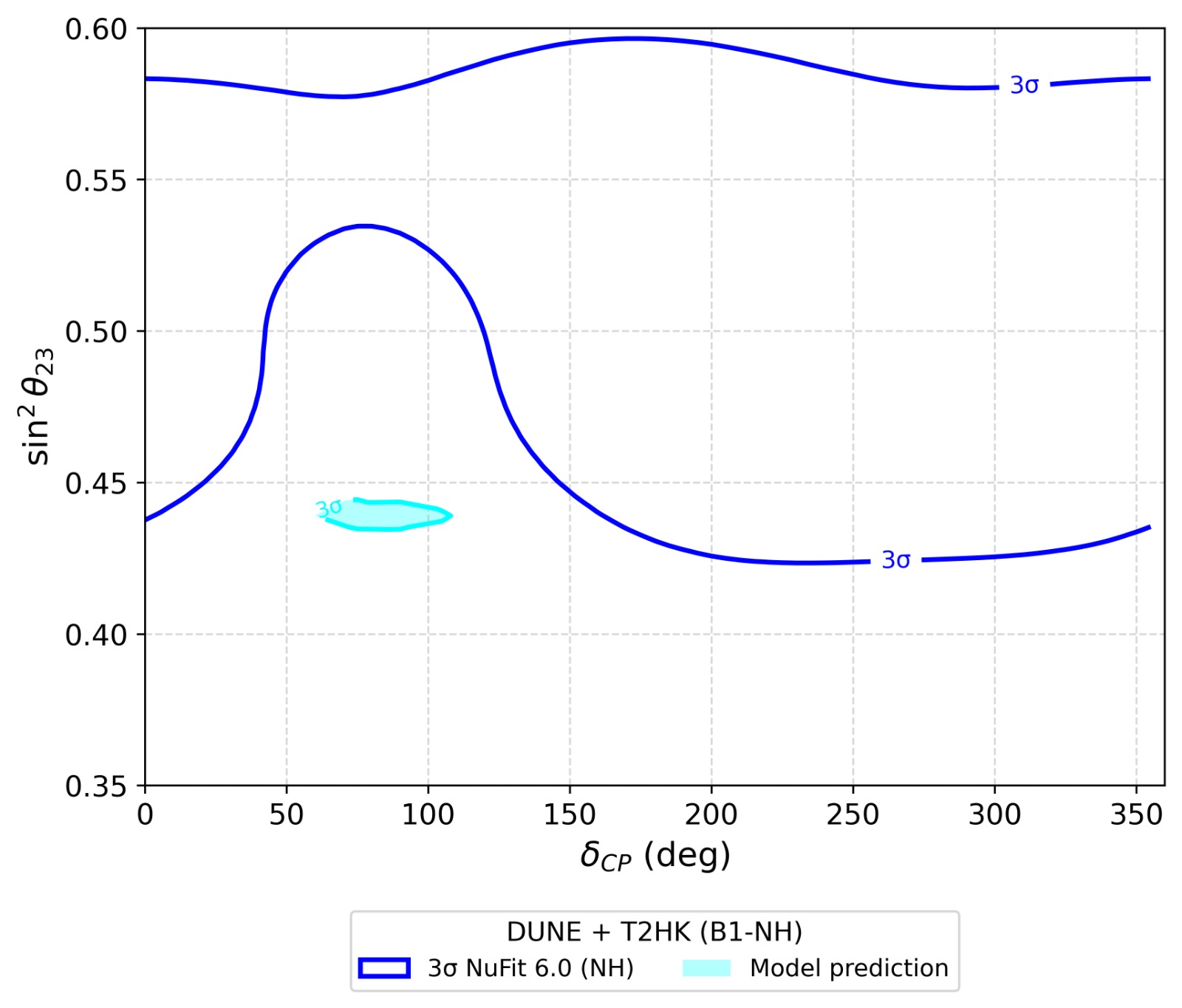}
    \caption{Similar $3\sigma$ contour like the main text for DUNE+T2HK in the $\sin^2\theta_{23}-\delta_{CP}$ plane, where the model prediction failed to satisfy the recent JUNO bound. }
    \label{fig:excluded}
\end{figure}

In Fig.~\ref{fig:excluded}, we present the allowed parameter space for the two-zero B1 texture in the NH scenario, which is already disfavored by the JUNO data. Although the B1 texture in the NH scenario is already disfavored by the JUNO data, a small allowed contour still appears in the DUNE+T2HK analysis around $\delta_{\rm CP}\sim (70^\circ$--$110^\circ)$ and $\sin^2\theta_{23}\sim 0.44$. However, this region lies entirely outside the $3\sigma$ {\tt NuFit 6.0} allowed NH region and is strongly constrained by the precise measurements from JUNO, particularly through the correlations involving $\theta_{12}$ and reactor observables. Therefore, after including priors on $\sin^2\theta_{12}$ and $\sin^2\theta_{13}$, the remaining allowed island disappears completely, leading to the exclusion of the B1-NH texture at DUNE+T2HK.

To understand the physical mechanism behind this exclusion, we can exploit the linear algebraic structure of the Majorana neutrino mass matrix elements.
When a Majorana mass matrix element is zero ($M^\nu_{ab} = 0$), the relation can be written in a compact, linear form based on the PMNS matrix elements ($U_{\alpha i}$):
\begin{equation}
    M^\nu_{ab} = \sum_{i=1}^3 U_{ai} U_{bi} \tilde{m}_i = 0
\end{equation}
where $\tilde{m}_1 = m_1$, $\tilde{m}_2 = m_2 e^{i2\alpha}$, and $\tilde{m}_3 = m_3 e^{i2(\zeta+\delta_{\text{cp}})}$ absorb the Majorana phases. For the $B_1$ texture with $M^\nu_{13} = 0$ and $M^\nu_{22} = 0$, we get two homogeneous linear systems for the complex masses:
\begin{equation}
 \begin{array}{cc}
U_{e1}U_{\tau 1}\tilde{m}_1 + U_{e2}U_{\tau 2}\tilde{m}_2 + U_{e3}U_{\tau 3}\tilde{m}_3 = 0\\U_{\mu 1}^2\tilde{m}_1 + U_{\mu 2}^2\tilde{m}_2 + U_{\mu 3}^2\tilde{m}_3 = 0      
 \end{array}   
\end{equation}

We solve these linear equations and completely eliminate the heavy atmospheric mass scale ($\tilde{m}_3$) and get:
\begin{equation}
    \frac{\tilde{m}_2}{\tilde{m}_1} = \frac{m_2}{m_1}e^{i2\alpha} = \frac{U_{e3}U_{\tau 3}U_{\mu 1}^2 - U_{e1}U_{\tau 1}U_{\mu 3}^2}{U_{e2}U_{\tau 2}U_{\mu 3}^2 - U_{e3}U_{\tau 3}U_{\mu 2}^2}e^{i2\alpha}
\end{equation}
Taking the absolute value of both sides gives a completely analytical expression for the mass ratio independent of the Majorana phases:
\begin{equation}
 \frac{m_2}{m_1} = \left| \frac{U_{e3}U_{\tau 3}U_{\mu 1}^2 - U_{e1}U_{\tau 1}U_{\mu 3}^2}{U_{e2}U_{\tau 2}U_{\mu 3}^2 - U_{e3}U_{\tau 3}U_{\mu 2}^2} \right|  \label{m21e} 
\end{equation}

JUNO's precision measurements dictate the best-fit value of solar mass splitting as $\Delta m_{21}^2 = 7.50 \times 10^{-5} \text{ eV}^2$. For any reasonable absolute mass scale $m_1 < 0.03 \text{ eV}$, the physical ratio must satisfy:
$$\frac{m_2}{m_1} = \sqrt{1 + \frac{\Delta m_{21}^2}{m_1^2}} \approx 1.0007$$

Interestingly, this $B_1$ texture structure is allowed for the IH mass pattern by both the {\tt NuFit 6.0} global-fit and the JUNO-expercted parameter space, whereas it is excluded for the NH mass ordering by the JUNO projection. We can justify all these from the analytical treatment as follows:
\begin{itemize}
    \item With the allowed range for the mixing angle and the $\delta_{CP}$ value from our analysis, this ratio from Eq. \eqref{m21e} falls below 1($\sim 0.87$), hence showing deviation from JUNO's prediction. 
    \item With the flexibility of the mixing angle from {\tt NuFit 6.0} global-fit results, for the same $\delta_{CP}$ value allowed within our analysis, the mass ratio from Eq. \eqref{m21e} smoothly falls within 1.3 to 1.4. 
    \item In the inverted hierarchy scenario, the physical mass splitting is governed by $m_1 = \sqrt{m_3^2 + |\Delta m_{31}^2|}$ \text{ and } $m_2 = \sqrt{m_3^2 + |\Delta m_{31}^2| + \Delta m_{21}^2}$. Because $m_2 > m_1$, the physical ratio $\frac{m_2}{m_1}$ must satisfy: $1.0 > \frac{m_2}{m_1} \ge 1.0523$. This relation holds good in our analysis for the maximal atmospheric mixing. 
\end{itemize}

Under {\tt NuFit 6.0}, the allowed region for $\sin^2\theta_{12}$ spans a wide $3\sigma$ horizontal belt ($0.275 \rightarrow 0.345$), and $\delta_{\text{CP}}$ is widely unconstrained. This allows the right-hand side expression of Eq. \eqref{m21e} to shift dynamically and intercept the physical ratio curve at a real, positive value for $m_1$ (specifically at $m_1 \approx 0.01\text{ eV}$). Once JUNO shrinks the $\sin^2\theta_{12}$ uncertainty by nearly half, it restricts the matrix combinations severely. The required value of $\theta_{12}$ needed to maintain a real-valued absolute mass scale $m_1$ is completely ruled out in JUNO's recent result.

One of the most important consequences of a texture being ruled out is related to the possible UV complete models which explain this texture. There have been several studies proposing flavor symmetric origins of texture-zeros as mentioned in Ref. [61-70] of the manuscript. For example, the newly ruled-out texture $B_1$ can be explained simply by incorporating Abelian discrete symmetries like $Z_4$ under which the lepton doublets and charged lepton singlets transform as 
\begin{equation}
    L_e (e_R) \rightarrow i L_e (e_R), L_\mu (\mu_R) \rightarrow -L_\mu (\mu_R), L_\tau \rightarrow -i L_\tau (\tau_R).
\end{equation}
Considering a type-II seesaw mechanism with three different $SU(2)_L$ triplet fields transforming as 
\begin{equation}
    \Delta_1 \rightarrow - \Delta_1, \Delta_2 \rightarrow -i \Delta_2, \Delta_3 \rightarrow i \Delta_3,
\end{equation}
we can explain the structure of the $B_1$ texture. There can be several different symmetry realizations of the same texture. As all of them predict the same texture-zero mass matrix together with diagonal charged lepton mass matrix, ruling out the texture with JUNO data has the power to rule out all these flavor symmetric interpretations.

\providecommand{\href}[2]{#2}\begingroup\raggedright\endgroup

\end{document}